\documentclass{lmcs}

\usepackage[english]{babel}
\usepackage[utf8]{inputenc}
\usepackage[T1]{fontenc}

\usepackage{mathtools}
\usepackage{savesym}
\usepackage{amsmath}
\savesymbol{openbox}
\usepackage{amsthm}
\usepackage{amssymb}
\usepackage{thmtools}
\usepackage{thm-restate}
\savesymbol{Bbbk}

\usepackage{xspace}
\usepackage[pdftex,noxy]{virginialake}
\vlnostructuressyntax

\usepackage{adjustbox}
\usepackage{proofzilla}
\usepackage{arydshln}
\usepackage{mathtools}
\usepackage{wasysym}

\usepackage{hyperref}
\usepackage{cleveref}

\theoremstyle{plain}
\newtheorem{theorem}{Theorem}[section]
\newtheorem{lemma}[theorem]{Lemma}
\newtheorem{proposition}[theorem]{Proposition}

\theoremstyle{definition}
\newtheorem{definition}[theorem]{Definition}
\newtheorem{example}[theorem]{Example}
\newtheorem{notation}[theorem]{Notation}

\theoremstyle{remark}
\newtheorem{remark}[theorem]{Remark}

\input{macros}

\commenter{old}[gray]
\commenter{matteo}[pzbrickred][m]
\commenter{willem}[pzblue][w]
\commenter{dominic}[pzgreen][d]

\begin{document}

\title[Canonical Proof Nets for IMLL]{
    Canonical Proof Nets 
    \\
    for Intuitionistic Multiplicative Linear Logic
}

\author[M. Acclavio]{Matteo Acclavio\lmcsorcid{0000-0002-0425-2825}}[a]
\author[W. Heijtjes]{Willem Heijtjes\lmcsorcid{0009-0001-8941-1150}}[b]
\author[D. Hughes]{Dominic J. D. Hughes\lmcsorcid{}}[c]

\address{FORM, University of Southern Denmark, Odense, Denmark}	

\address{University of Bath, Bath, UK}	

\address{Logic Group, University of California Berkeley, CA, USA}	




\begin{abstract} 
 	We study the problem of finding canonical representations of proofs in the multiplicative fragment of intuitionistic linear logic with units.

    We show that, contrary to the classical case, it is possible to define a notion of proof nets for this fragment which is provided with a polynomial-time sequentialization procedure and is both canonical and efficient.
    In particular, we show that sequent calculus derivations differing only by a permutation of independent rule applications are translated to the same proof net, and that the translations between sequent calculus and proof nets are inverse to each other.
\end{abstract}

\maketitle

\section{Introduction}\label{sec:intro}

\emph{Proof nets} are a graph-based syntax for proofs introduced by Girard \cite{girard:PN} in the context of linear logic \cite{gir:ll}.
One of its motivations was to provide an analogous of natural deduction for linear logic, that is, a syntax for proofs that would be free of the bureaucratic aspects of the sequent calculus, such as the order of independent rule applications.
The central idea of proof nets is to replace the sequent calculus rules with  graph-based constructs, called \emph{gates}, connecting active and principal formulas of rules, and to represent proofs as graphs generated by these gates.
The benefit in replacing sequent calculus rules with gates is the fact that they make it possible to track dependencies between rule applications, while abstracting away from the ordering of rules that operate on disjoint sets of active and principal formulas.

\begin{equation}\label{eq:introDerToPNold}
    \adjustbox{max width = \textwidth}{$
    \begin{array}{c}
        \vlderivation{
            \vliin{\ltens}{}{\vdash \lneg a, a \ltens \lneg b, b\ltens\lneg c, c}{
                \vliin{\ltens}{}{
                    \vdash \lneg a, a \ltens \lneg b, b
                }{
                    \vlin{\axrule}{}{\vdash \lneg a, a}{\vlhy{}}    
                }{
                    \vlin{\axrule}{}{\vdash \lneg b, b}{\vlhy{}}
                }
            }{
                \vlin{\axrule}{}{\vdash \lneg c, c}{\vlhy{}}
            }
        }
    \qquad\preq\qquad
        \vlderivation{
            \vliin{\ltens}{}{\vdash \lneg a, a \ltens \lneg b, b\ltens\lneg c, c}{
                \vlin{\axrule}{}{\vdash \lneg a, a}{\vlhy{}}    
            }{
                \vliin{\ltens}{}{
                    \vdash \lneg b, b \ltens \lneg c,c
                }{
                    \vlin{\axrule}{}{\vdash \lneg b, b}{\vlhy{}}
                }{
                    \vlin{\axrule}{}{\vdash \lneg c, c}{\vlhy{}}
                }
            }
        }
    \\[40pt]
        \begin{array}{ccccccccc}
            &\vgate{axa}{\axrule} && \vgate{axb}{\axrule}  && \vgate{axc}{\axrule} 
            \\[10pt]
            &&\vgate{L}{\ltens} &&\vgate{R}{\ltens}
            \\[10pt]
            \vpz1{\lneg a} && \vpz2{a \ltens \lneg b} && \vpz3{ b \ltens \lneg c} && \;\;\vpz4{c}
        \end{array}
        \loosePNedges{
            gateaxa/pz1/1.5/180/90,
            gateaxa/gateL.140/1.5/0/90,
            gateaxb/gateL.40/1.5/180/90,
            gateaxb/gateR.140/1.5/0/90,
            gateaxc/gateR.40/1.5/180/90,
            gateaxc/pz4/1.5/0/90
        }
        \PNedges{gateL/pz2,gateR/pz3}
    \end{array}
    $}
\end{equation}

In the multiplicative fragment of linear logic ($\MLL$), the tree-like structure of formulas already captures all the dependencies between rules, and thus only missing information to fully identify a proof is the pairing of dual formulas corresponding to the logical axioms used in the derivation.
\begin{equation}\label{eq:introPNtoPNnew}
    \vpz{na}{\lneg a} 
    \quad 
    \vpz{a}{a} \ltens \vpz{nb}{\lneg b} 
    \quad 
    \vpz{b}{b} \ltens \vpz{nc}{\lneg c} 
    \quad 
    \vpz{c}{c}
    \sqBedges{pzna/pza/12,pznb/pzb/12,pznc/pzc/12}
\end{equation}
This yields canonical representations of proofs: sequent calculus derivations that differ only in the order of independent rule applications are represented by the same proof net.%
\footnote{
    It is worth noting that no proof of cut-elimination in the literature can avoid dealing with independent rule permutations, since cut-elimination steps are typically defined only for the case where both active formulas of the cut are principal for the rules immediately above it. Therefore, so-called \emph{commutative steps} are required to permute cuts with other rules.
}

However, trade-offs are inevitable. 
While proof nets eliminate the bureaucracy of sequent calculus syntax, they introduce the need for \emph{correctness criteria} to distinguish \emph{proof structures} (i.e., graphs freely constructed using gates) from \emph{proof nets} (i.e., graphs encoding correct proofs). 

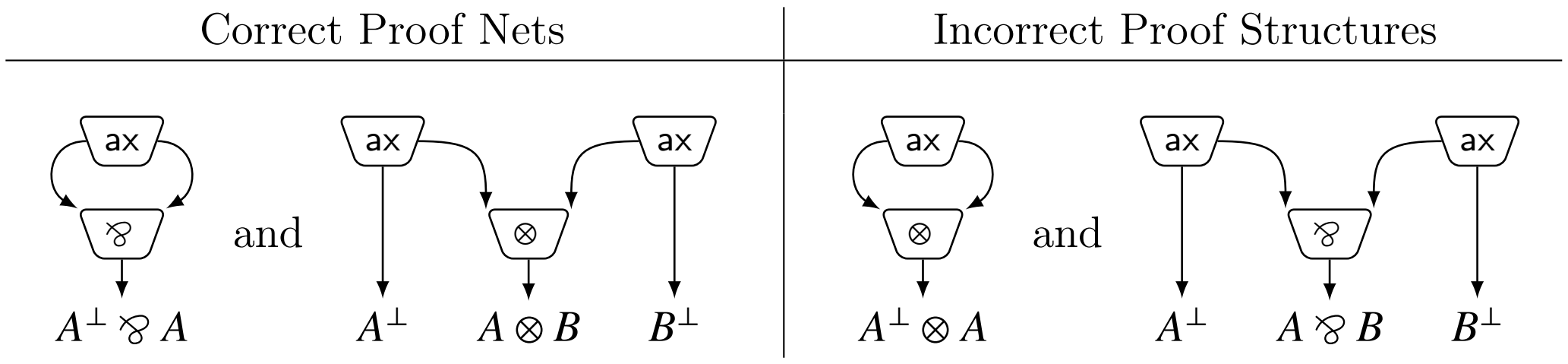
\begin{equation}\label{eq:introCorrect}
    \begin{array}{c@{\quad\vrule\quad}c}
        \text{Correct Proof Nets}
        &
        \text{Incorrect Proof Structures}
    \\\hline\\
        \begin{array}{c}
            \vgate1{\axrule} 
            \\[10pt]
            \vgate0{\lpar} 
            \\[10pt]
            \vpz1{\lneg A \lpar A}
        \end{array}
        \loosePNedges{
            gate1/gate0.30/1.5/0/30,
            gate1.180/gate0.150/1.5/180/150
        }
        \PNedges{gate0/pz1}
    \text{ and }
        \begin{array}{ccc}
            \vgate{axL}{\axrule} & &\vgate{axR}{\axrule} 
            \\[10pt]
            &\vgate0{\ltens} 
            \\[10pt]
            \vpz1{\lneg A }&\vpz2{ A \ltens B}& \vpz3{\lneg B}
        \end{array}
        \loosePNedges{
            gateaxL/gate0.150/1.5/0/90,
            gateaxR.180/gate0.30/1.5/180/90
        }
        \PNedges{gateaxL/pz1,gate0/pz2,gateaxR/pz3}
    & 
        \begin{array}{c}
            \vgate1{\axrule} 
            \\[10pt]
            \vgate0{\ltens} 
            \\[10pt]
            \vpz1{\lneg A \ltens A}
        \end{array}
        \loosePNedges{
            gate1/gate0.30/1.5/0/30,
            gate1.180/gate0.150/1.5/180/150
        }
        \PNedges{gate0/pz1}
    \text{ and }
        \begin{array}{ccc}
            \vgate{axL}{\axrule} & &\vgate{axR}{\axrule} 
            \\[10pt]
            &\vgate0{\lpar} 
            \\[10pt]
            \vpz1{\lneg A }&\vpz2{ A \lpar B}& \vpz3{\lneg B}
        \end{array}
        \loosePNedges{
            gateaxL/gate0.150/1.5/0/90,
            gateaxR.180/gate0.30/1.5/180/90
        }
        \PNedges{gateaxL/pz1,gate0/pz2,gateaxR/pz3}
    \end{array}
\end{equation}

Moreover, although the basic idea of proof nets works elegantly for the multiplicative fragment of linear logic, it does not generalize well in the presence of rules with a structural character, such as weakening and contraction, or even units and additive connectives.
The root of this issue lies in the fact that the rules-as-gates approach fail to capture non-local dependencies.
For example, the rule for the negative multiplicative unit $\lbot$ (see \Cref{fig:rules}) can only be applied to a non-empty context, while the rule for additive conjunction requires that both premises share the same context.%
\footnote{
    More recent approaches to proof nets, such as slice nets \cite{hug:van:slice,hug:van:sliceLICS}, conflict nets \cite{hughes:conflict,hei:hug:conflict} for $\MALL$, and expansion trees for second-order $\MLL$ \cite{str:MLL2}, but also combinatorial proofs \cite{hughes:invar,hughes:pws,acc:str:IJCAR18,hughes:fopws,ICP,acc:str:CPK,acc:EHPN,acc:str:AIML22} abandon the idea of using gates altogether.
}

\begin{figure}
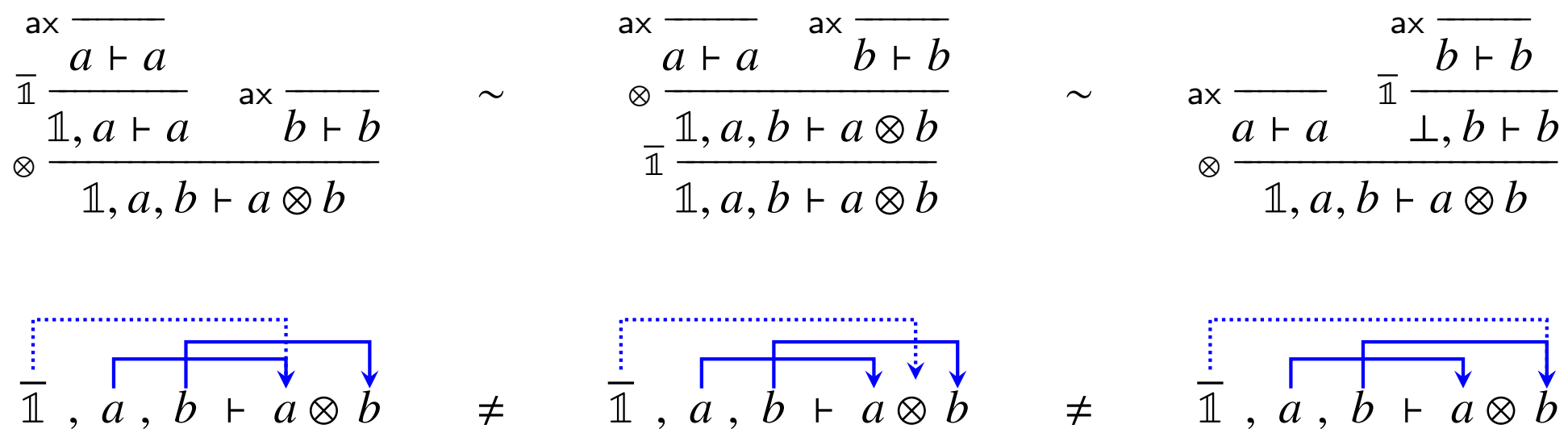

    \def\myskip{\hskip2em}
    $$
    \begin{array}{c@{\myskip}c@{\myskip}c@{\myskip}c@{\myskip}c}
        \vlderivation{
            \vliin{\rltens}{}{ \lone, a, b\vdash a\ltens b}{
                \vlin{\lonel}{}{\lone ,a \vdash  a}{\vlin{\axrule}{}{a \vdash a}{\vlhy{}}}
            }{
                \vlin{\axrule}{}{b \vdash b}{\vlhy{}}
            }
        }
    &\preq&
        \vlderivation{
            \vlin{\lonel}{}{
                \lone ,a,b \vdash  a\ltens b
            }{
                \vliin{\rltens}{}{ \lone, a, b\vdash a\ltens b}{
                    \vlin{\axrule}{}{a \vdash a}{\vlhy{}}
                }{
                    \vlin{\axrule}{}{b \vdash b}{\vlhy{}}
                }
            }
        }
    &\preq&
        \vlderivation{
            \vliin{\rltens}{}{\lone, a, b\vdash a\ltens b}{
                \vlin{\axrule}{}{a \vdash  a}{\vlhy{}}
            }{
                \vlin{\lonel}{}{\lbot ,b \vdash  b }{\vlin{\axrule}{}{b \vdash  b}{\vlhy{}}}
            }
        }
    \\\\[20pt]
        \vpz1{\none} \;,\; \va1 \;,\; \vb1 \;\vdash\; \va0\vpz0{\ltens} \vb0
        \sqDBedges{a1/a0/12,b1/b0/16}
        \sqdDBedges{pz1/a0/20}
    &\neq&
        \vpz1{\none} \;,\; \va1 \;,\; \vb1 \;\vdash\; \va0\vpz0{\ltens} \vb0
        \sqDBedges{a1/a0/12,b1/b0/16}
        \sqdDBedges{pz1/pz0/20}
    &\neq&
        \vpz1{\none} \;,\; \va1 \;,\; \vb1 \;\vdash\; \va0\vpz0{\ltens} \vb0
        \sqDBedges{a1/a0/12,b1/b0/16}
        \sqdDBedges{pz1/b0/20}
    \end{array}
    $$

    \caption{
        Three distinct derivations equivalent derivations and their corresponding proof nets.
        The one in the middle is the \emph{canonical proof net} providing the canonical representative of equivalence class.
    }
    \label{fig:introExample}
\end{figure}

\myparagraph{The problem of multiplicative units}
The design of a syntax for proof nets that allows for canonical representations of proofs in $\MLL$ with units admitting both a polynomial-time correctness criterion a polynomial-time proof translation has been a long-standing open problem.
A key idea to recover a way to easily check the correctness of an application of the $\lbot$-rule is to include additional edges connecting a $\lbot$-gate to another gate in the proof net, called \emph{jumps}, to represent the connection between the $\lbot$-gate and the context in which it is applied.
\footnote{
    Some works such as \cite{dig:fag:jumps} have exploited this idea of reintroducing additional dependencies between rules in proof nets using jumps, beyond the strict necessity to ensure correctness.
}
This makes it possible to ensure that the $\lbot$-gate corresponds to a valid application of the $\lbot$-rule, even in the absence of syntactic structure indicating the entire surrounding context.
However, while jumps guarantee polynomial-time proof translation and correctness criterion, the presence of jumps breaks canonicity: 
two derivations that should be considered equivalent modulo rule permutations may be represented by different proof nets solely because of jumps, as shown in \Cref{fig:introExample}.

The work of Heijltjes and Houston \cite{heijltjes:houston:14} ruled out the existence of a satisfactory solution to this problem capable of balancing canonicity with efficiency of both proof translation and correctness criterion%
\footnote{
    As a consequence, this also applies to proof nets for the multiplicative exponential fragment of linear logic ($\MELL$), where the weakening rule cause similar non-canonicity issues.
} 
Their result follows from showing that the reconfiguration problem \cite{reconfiguration} in constraint logic \cite{constraintLogic} can be reduced to the problem of proof equivalence in the fragment of $\MLL$ containing only units dealing with jumps rewiring.%
\footnote{
    It is worthy to remind the reader that simply removing all jumps in not a solution: dropping the connectedness condition from the correctness criterion would correspond to consider the rule $\mixr$, and that weakening connectedness, as done in \cite{regnier:phd,did:tor:25}, would only isolate ad-hoc fragments.
    On the other hand, one could naively translate a proof into the set of all `equivalent' proof nets with jumps, but this would lead to an exponential blow-up in the size of the proof net, as the one encountered in monomial and Slice nets for $\MALL$ \cite{girard:PN,hug:van:slice,hug:van:sliceLICS}.
}
Thus, if canonical proof nets with a polynomial-time proof translation existed, we could translate two derivations into proof nets and test their equivalence in polynomial time, implying that the complexity classes $\comPtime$ and $\comPspace$ should coincide.
Of course, one could `cheat' and define a canonical proof net by defining it as the set of all possible proof nets with jumps corresponding to a given derivation, but this would lead to an exponential blow-up in the size of the proof net, as the one encountered in monomial and Slice nets for $\MALL$ \cite{hug:van:slice,hug:van:sliceLICS}.

Therefore, under the assumption $\comPtime\neq\comPspace$, we have the following consequence:
\begin{quote}
    \centering
    \emph{
        There cannot be \textbf{canonical} proof nets for $\MLL$ (with units) 
        with $\comPtime$ proof translation and $\comPtime$ correctness criterion.
    }
\end{quote}

At the time of writing, it is still an open question whether the multiplicative fragment of intuitionistic linear logic ($\IMLL$) admits proof nets with a polynomial-time correctness criterion and polynomial-time proof translation capturing the equivalence induced by independent rule permutations.
If one considers the negative results for $\MLL$ with units, one could be tempted to conclude that the same result should hold for $\IMLL$, but this is not the case. 
However, Lamarche's correctness criterion for essential nets \cite{lamarche:essential} restricted to $\IMLL$, while similar to Danos-Regnier switching criterion for $\MLL$ \cite{danos:regnier:89}, is simpler and more efficient, leaving room for a possible solution to the problem of canonicity for $\IMLL$.

\paragraph*{Contribution.}
In this paper, we show that the problem of canonicity for $\IMLL$ admits a quite simple solution, and that the problem of canonicity for $\MLL$ does not affect $\IMLL$.
For this, we look at Lamarche's proof nets for $\IMLL$ \cite{lamarche:essential}, and we show that there is a simple natural way to exploit the same mechanism used to check correctness to define a canonical way to assign jumps to negative units.

The idea is to observe that the correctness criterion for proof nets for $\IMLL$ naturally defines paths oriented towards the consequent of the sequent, and that we can use these paths to define a canonical way to assign jumps to negative units.
For example, while the three derivations in \Cref{fig:introExample} are all equivalent, their corresponding proof nets differ in the way jumps are assigned to the negative unit $\lunit$.
However, the proof net in the middle is the only one which is \emph{canonical}, because the jumps cannot be pushed forward along the paths defined by the correctness criterion.

\begin{figure}[t]
    $$
    \begin{tikzpicture}
        \draw[draw, rounded corners=10] (-1,0) rectangle (7,2);
        \draw[draw, rounded corners=10] (3,0) rectangle (11,2);
        \draw[draw, rounded corners=10] (1,1) rectangle (9,-1);
        \node at (1,1.5) {Sequentializable};
        \node at (9,1.5) {Total};
        \node at (5,1.5) {Proof Net};
        \node at (5,-.5) {Maximal};
        \node at (5,.5) {Canonical Net};
    \end{tikzpicture}
    $$

    \caption{
        Relations between different properties of proof structures for $\IMLL$ with units.
        Here sequentializable means there is a polynomial time procedure to construct a derivation in the sequent calculus, total means that each negative unit has an associated jump, and maximal means that the jumps are assigned `as closer to the output as possible' to the consequent of the sequent with respect to the order induced by the correctness criterion.
    } 
    \label{fig:venn}
\end{figure}
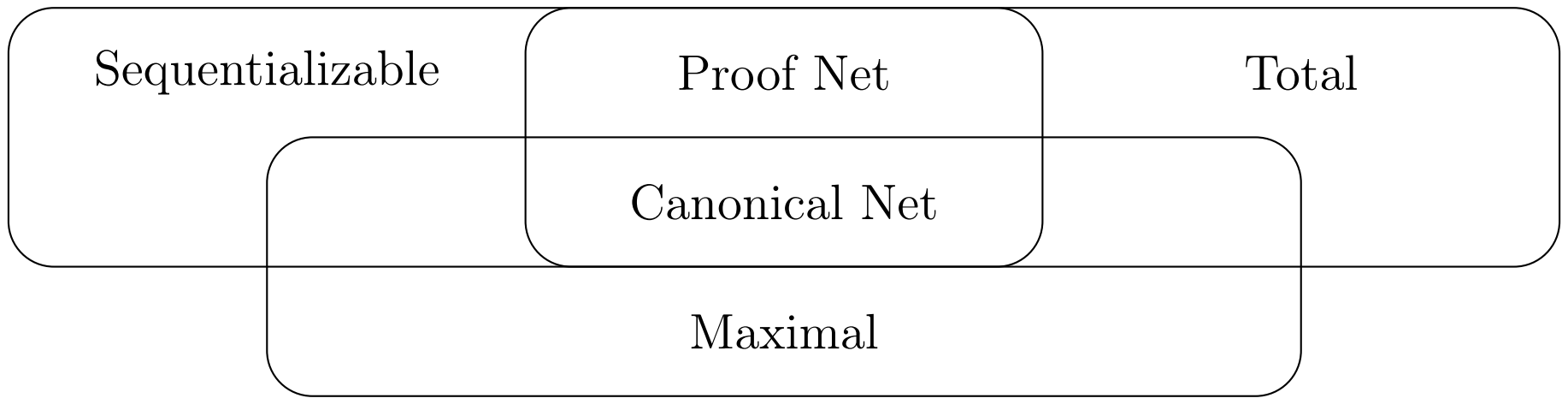

The properties on the resulting proof nets are summarized in \Cref{fig:venn}.
Interestingly, we can show a surprising result which does not hold for $\MLL$: it is possible to define a polynomial-time correctness criterion and sequentialization procedure for proof nets without jumps which is sound and complete for $\IMLL$ (with units).
Similar attempt to define proof nets for $\IMLL$ without jumps have been proposed in the literature, but they all fail to be complete.

\paragraph*{Structure of the paper.}
In \Cref{sec:back} we recall the definition of the sequent calculus for $\IMLL$ and its proof nets, and we fix the notation for formulas, sequents, and formula trees we use in the paper.
In \Cref{sec:pn} we recall the definition of proof nets for $\IMLL$ and we recall Lamarche's correctness criterion for pure multiplicative essential nets.
In \Cref{sec:canon} we define a canonical way to rewire jumps from negative units to positive formulas, and we show that the resulting proof nets are canonical, and we define a cut-elimination procedure for canonical proof nets.
In \Cref{sec:jumpless} we discuss a correctness criterion for proof nets without jumps enforcing a stronger quotient on proofs.
We conclude in \Cref{sec:conc} with some remarks and future directions.

\section{Background}\label{sec:back}

In this section we briefly fix the notation and recall the definition of the sequent calculus for $\IMLL$ and its proof nets, which we will use in the rest of the paper.

\subsection{Intuitionistic Multiplicative Linear Logic}\label{subsec:IMLL}

\defn{Formulas} are  generated from a countable set of propositional variables (or \defn{atoms}) $\atoms= \set{a, b, \dots}$ and the \defn{unit} ($\lone$) using the logic operators \defn{linear implication} ($\limp$), and \defn{tensor} ($\ltens$) as follows:
\begin{equation}\label{eq:formulas}
    \hfill
    \begin{array}{l@{\quad}c@{\quad\vrule\quad}c@{\quad\vrule\quad}c@{\quad\vrule\quad}c}
        A,B \;\coloneqq &a & \lone &  A\limp B & A\ltens B
    \end{array}
    \hfill
\end{equation}

A \defn{context} $\Gamma = A_1, \dots, A_n$ is a comma-separated list of formulas sequence which we interpret as a multiset, that is, the order of formulas in a context does not matter, and formulas can occur multiple times in a context.
A \defn{sequent} $\Gamma \vdash A$ comprises a context $\Gamma$ (the \defn{antecedent}) and a formula $A$ (the \defn{consequent}).

In this paper, we identify formulas with their parse trees, thus occurrences of atoms, units, and connectives in a formula are identified with the unique corresponding node of the parse tree.
For example, we identify the formula 
\begin{equation}\label{eq:sequent}
    \Big(\big(((p\ltens\lone)\limp p)\limp (q\ltens\lone)\big)\ltens( \lone \limp \lone) \Big)\limp ((p\limp p )\ltens q)
\end{equation}
with its parse tree, which is depicted as follows:
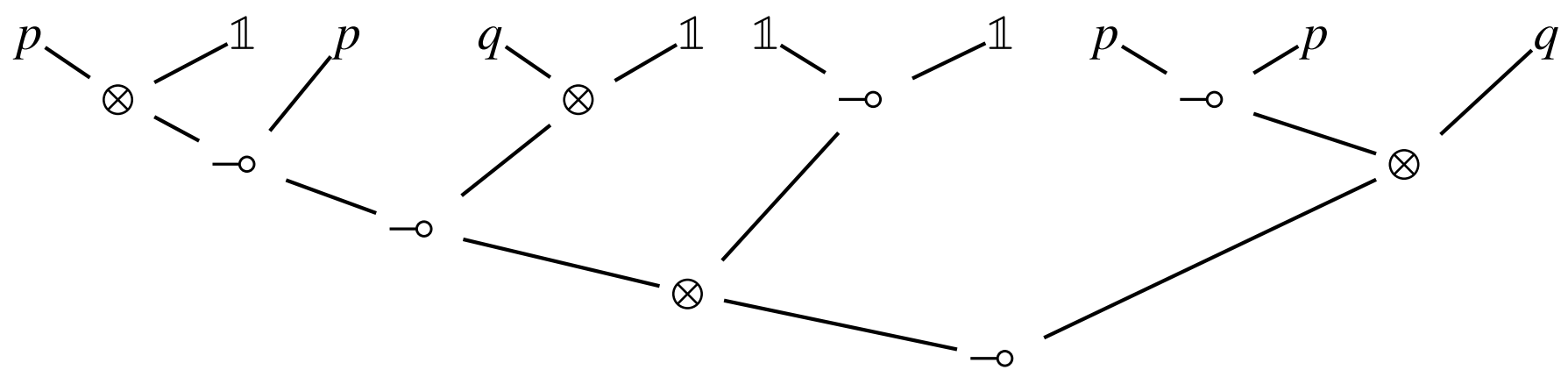
\begin{equation}\label{eq:parse-tree}
    \begin{array}{cccc@{\;}c@{\;}ccc@{\;}c@{\;}ccccccccccc}
        \vp1&& \vlone1 & \vp4 && \vq1 && \vlone 2 && \vlone3 && \vlone 4 & \vp2 && \vp3 &&& \vq2
    \\
        &\vpz1{\ltens}
        &&&&
        &\vpz2{\ltens}
        &&&
        &\vpz3{\limp}
        &&&
        \vpz4{\limp}
    \\
        &&
        \vpz5{\limp}
        &&&&&&
        &&&&&&&
        \vpz6{\ltens}
    \\
        &&&&\vpz7{\limp}
    \\
        &&&&&&&\vpz9{\ltens}
    \\
        &&&&&&&&&&&\vpz8{\limp}
    \end{array}
    \Gedges{p1/pz1,lone1/pz1,pz1/pz5,q1/pz2,p4/pz5,lone2/pz2,pz5/pz7,pz2/pz7,pz3/pz9}
    \Gedges{lone3/pz3,lone4/pz3,p2/pz4,p3/pz4,pz4/pz6,q2/pz6,pz9/pz8,pz6/pz8,pz9/pz7}    
\end{equation}

A \defn{node} of a sequent is any node of the parse tree of a formula occurring in a sequent, that is, any occurrence of an atom, unit, or connective in a formula occurring in a sequent.
Two sub-formulas in a sequent are \defn{disjoint} if they do not share any node.

\begin{definition}
    The \defn{implication depth} of a sequent node is the number of times it occurs in the left  argument (sub)formula of a $\limp$ or $\vdash$. 
    For example, each node in the following sequent has its implication depth annotated over it:
    $$
    \begin{array}{c@{}ccc@{}c@{\;}c@{}c@{}ccc@{}ccccccccccccccc}
        ((& p & \ltens & \lone &)\limp& p &)\limp (& q &\ltens& \lone & ),& \lone &\limp& \lone &\vdash (&p &\limp& p&) \ltens &q
    \\
        &3 & 3 & 3 & 2 & 2 & 1 & 1 & 1 & 1 && 2 & 1 & 1 && 1 & 0 & 0 & 0 & 0
    \end{array}
    $$

    We say that a node is \defn{positive} if its implication depth is even, and \defn{negative} if its implication depth is odd.
\end{definition}

\begin{notation}\label{notation:overline}
    In the following, we may annotate nodes with negative variance by decorating them with a overline.
    For example, we may write the sequent above as follows:
    $$
    (( \cneg p  \ntens  \none )\limp p ) \nimp ( \cneg q \ntens \none  ), \lone \nimp \none \vdash (\cneg p \limp p) \ltens q
    $$
    Similarly, if a sub-formula has its root node with negative variance, we may write it with a overline, for example:
    $$
    (\overline{( p  \ltens  \lone )}\limp p ) \limp \overline{(  q \ltens \lone  )}, \overline{\lone \limp \lone }\vdash (\overline p \limp p) \ltens q
    \qdot
    $$
\end{notation}

In \Cref{fig:rules} recall the definition of the sequent calculus for \defn{intuitionistic multiplicative linear logic} ($\IMLL$), and we recall the admissible rules for the generalized axiom and cut rules, which we will use in the rest of the paper.
We say that a formula is \defn{active} \resp{\defn{principal}} for a rule if it occurs in a premise \resp{in the conclusion} of the rule but not in the conclusion \resp{in none of the premises}.

\begin{remark}
    Notice that for the rules for the implication and tensor connectives we adopt names which reflect whether the main connective of the active formula negative or positive, that is, if it occurs in the antecedent or in the consequent of the sequent respectively.
\end{remark}

A \defn{derivation} is a tree of sequents such that each node of the tree is the conclusion of a rule instance whose premises are the children of the node. In particular, its leaves are conclusion of the rules $\axrule$ and $\lone$.

\begin{theorem}
    In $\IMLL$, the following hold:
    \begin{itemize}
        \item the \emph{general axiom} rule $\AXrule$ is derivable;
        \item the rule $\cutr$ is admissible.
    \end{itemize}
\end{theorem}
\begin{proof}
    To prove the derivability of $\AXrule$, we proceed by induction on $A$.
    If $A$ is an atom, then the derivation is just an instance of $\axrule$.
    If $A$ is $\lone$, then the derivation is an instance of $\none$ followed (bottom-up) by an instance of $\lone$.
    If $A$ is $B\limp C$ \resp{$B\ltens C$}, then the derivation is an instance of $\limp$ \resp{$\ntens$} followed (bottom-up) by a $\nimp$ \resp{$\ltens$} with premises the sequents $B\vdash B$ and $C\vdash C$, which are both derivable by induction hypothesis.

    To prove the admissibility of $\cutr$, we use the \emph{cut-elimination steps} recalled in \Cref{fig:cutelim} to push $\cutr$-rules upwards in the derivation until they reach the axioms, where they can be eliminated.
    Termination of the cut-elimination procedure follows from the fact that 
    each principal cut-elimination step strictly decreases the size of the cut formula, while 
    each commutative cut-elimination step strictly decreases the maximum length of the branches above the $\cutr$-rule in the derivation.
\end{proof}

\begin{figure}[!t]
    \centering
	\def\myskip{\hskip1.5em}
	\adjustbox{max width=\textwidth}{$\begin{array}{cccc|c}
        \vlinf{\axrule}{}{a \vdash a}{}
    &
        \vlinf{\loner}{}{\vdash \lone}{}
    &
        \vliinf{\ltens}{}{ \Gamma, \Delta\vdash A\ltens B}{\Gamma \vdash A}{\Delta \vdash B}
    &
        \vliinf{\nimp}{}{ \Gamma, A\limp B, \Delta \vdash C}{\Gamma
        \vdash A}{B,\Delta\vdash C}
    &
        \vliinf{\cutr}{}{\Gamma, \Delta \vdash  A}{ \Gamma\vdash C}{C,\Delta\vdash A}
    \\
    &
        \vlinf{\lonel}{}{\Gamma,\lone\vdash A}{\Gamma\vdash A}
    &
        \vlinf{\ntens}{}{\Gamma, A\ltens B \vdash C}{ \Gamma, A, B \vdash C}
    &
        \vlinf{\limp}{}{\Gamma \vdash A\limp B}{ \Gamma, A\vdash B}
    &
        \vlinf{\AXrule}{}{A \vdash A}{}
    \end{array}$}
	\caption{
        Sequent rules for $\IMLL$, and the rules for the generalized $\AXrule$ and $\cutr$.
    }
	\label{fig:rules}
\end{figure}

\begin{figure}[!t]
    \centering
	\adjustbox{max width=\textwidth}{$\begin{array}{ccc}
    \\
        \vlderivation{
            \vliin{\cutr}{}{
                \Gamma\vdash A
            }{
                \vlin{\none}{}{\Gamma,\lone\vdash A}{\vlhy{\Gamma\vdash A}}
            }{
                \vlin{\lone}{}{\vdash \lone}{\vlhy{}}
            }
        }
    \quad\cutelim\quad
        \Gamma\vdash A
    &&
        \vlderivation{
            \vliin{\cutr}{}{
                \Gamma\vdash B
            }{
                \vlin{\axrule}{}{a\vdash a}{\vlhy{}}
            }{
                \vlhy{\Gamma, a \vdash B}
            }
        }
    \quad\cutelim\quad
        \Gamma\vdash B
    \\\\
        \vlderivation{
            \vliin{\cutr}{}{
                \Gamma,\Delta,\Sigma\vdash C
            }{
                \vlin{\rlimp}{}{\Gamma \vdash A\limp B}{\vlhy{\Sigma, A\vdash B}}
            }{
                \vliin{\llimp}{}{
                    \Delta, A\limp B, \Sigma \vdash C
                }{
                    \vlhy{\Delta\vdash A}
                }{
                    \vlhy{B,\Sigma\vdash C}
                }
            }
        }
    &\quad\cutelim\quad&
        \vlderivation{
            \vliin{\cutr}{}{
                \Gamma,\Delta,\Sigma\vdash C
            }{
                \vliin{\cutr}{}{\Delta,\Sigma \vdash B}{\vlhy{\Delta\vdash A}}{\vlhy{\Sigma, A\vdash B}}
            }{
                \vlhy{B,\Sigma\vdash C}
            }
        }
    \\\\
        \vlderivation{
            \vliin{\cutr}{}{
                \Gamma,\Delta\vdash C
            }{
                \vliin{\rltens}{}{ \Gamma, \Delta\vdash A\ltens B}{\vlhy{\Gamma \vdash A}}{\vlhy{\Delta \vdash B}}
            }{
                \vlin{\lltens}{}{\Sigma, A\ltens B \vdash C}{\vlhy{\Sigma, A, B \vdash C}}
            }
        }
    &\quad\cutelim\quad&
        \vlderivation{
            \vliin{\cutr}{}{
                \Gamma,\Delta\vdash C
            }{
                \vlhy{\Gamma \vdash A}
            }{
                \vliin{\cutr}{}{ \Delta, \Sigma,A \vdash C}{\vlhy{\Delta \vdash B}}{\vlhy{\Sigma, A, B \vdash C}}
            }
        }
    \\\\\hline\\
    \\[5pt]
        \vlderivation{
            \vliin{\cutr}{}{
                \Gamma,\Delta \vdash C
            }{
                \vlin{\set{\lonel,\ntens}}{}{
                    \Gamma\vdash A
                }{
                    \vlhy{\Gamma'\vdash A}
                }
            }{
                \vlhy{A, \Delta \vdash C}
            }
        }
    &\quad\cutelim\quad&
        \vlderivation{
            \vlin{\set{\lonel,\ntens}}{}{
                \Gamma, \Delta \vdash C
            }{
                \vliin{\cutr}{}{
                    \Gamma',\Delta \vdash C
                }{
                    \vlhy{\Gamma'\vdash A}
                }{
                    \vlhy{A, \Delta \vdash C}
                }
            }
        }
    \\\\
        \vlderivation{
            \vliin{\cutr}{}{
                \Gamma,\Delta \vdash C
            }{
                \vlhy{\Gamma \vdash A}
            }{
                \vlin{\set{\lonel,\limp,\ntens}}{}{
                    A,\Delta \vdash C
                }{
                    \vlhy{A,\Delta'\vdash C'}
                }
            }
        }
    &\quad\cutelim\quad&
        \vlderivation{
            \vlin{\set{\lonel,\limp,\ntens}}{}{
                \Gamma, \Delta \vdash C
            }{
                \vliin{\cutr}{}{
                    \Gamma,\Delta' \vdash C'
                }{
                    \vlhy{\Gamma\vdash A}
                }{
                    \vlhy{A, \Delta' \vdash C'}
                }
            }
        }
    \\\\
        \vlderivation{
            \vliin{\cutr}{}{
                \Gamma,B\limp D, \Delta,\Sigma \vdash C
            }{
                \vliin{\nimp}{}{
                    \Gamma,B\limp D, \Delta\vdash A
                }{
                    \vlhy{\Gamma\vdash B}
                }{
                    \vlhy{D, \Delta \vdash A}
                }
            }{
                \vlhy{A, \Sigma \vdash C}
            }
        }
    &\quad\cutelim\quad&
        \vlderivation{
            \vliin{\nimp}{}{
                \Gamma,B\limp D, \Delta,\Sigma \vdash C
            }{
                \vlhy{\Gamma \vdash B}
            }{
                \vliin{\cutr}{}{
                    D,\Delta,\Sigma \vdash A
                }{
                    \vlhy{D,\Delta \vdash A}
                }{
                    \vlhy{A, \Sigma \vdash C}
                }
            }
        }
    \\\\
        \vlderivation{
            \vliin{\cutr}{}{
                \Gamma,\Delta, \Sigma \vdash B\ltens C
            }{
                \vlhy{\Gamma \vdash A}
            }{
                \vliin{\ltens}{}{
                    A, \Delta, \Sigma \vdash B\ltens C
                }{
                    \vlhy{A, \Delta \vdash B}
                }{
                    \vlhy{\Sigma \vdash C}
                }
            }
        }
    &\quad\cutelim\quad&
        \vlderivation{
            \vliin{\ltens}{}{
                \Gamma,\Delta, \Sigma \vdash B\ltens C
            }{
                \vliin{\cutr}{}{
                    \Gamma, \Delta \vdash B
                }{
                    \vlhy{\Gamma \vdash A}
                }{
                    \vlhy{A, \Delta \vdash B}
                }
            }{
                \vlhy{\Sigma \vdash C}
            }
        }
    \end{array}$}

	\caption{
        Cut-elimination steps for $\IMLL$.
    }
	\label{fig:cutelim}
\end{figure}

\subsection{Graphs}\label{subsec:GandT}

We recall standard definitions for graphs and formula trees to fix the notation we use in the paper.
Then, we define a natural notion of \emph{\variance} for nodes of the forest encoding the formulas in a sequent which plays a crucial role in the definition of our proof nets for $\IMLL$ with units.

A \defn{(direct) graph} defined as triples of the form $\gG=\tuple{\vof[\gG],\dof[\gG]}$ made of 
a (finite) set of \defn{vertices} $\vof[\gG]$, 
and 
a set of \defn{edges} $\dof[\gG]$ whose elements are pairs of vertices, written $vw$ or $v\dof w$.
In an edge $v\dof w$, we call $v$ its \defn{source} $v$ and $w$ its \defn{target}. 
Two vertices are \defn{adjacent} if there is an edge connecting them.
The \defn{size} of a graph is the sum of the number of its vertices and edges.

A \defn{path} in a graph $\gG$ form a vertex $u$ to a vertex $w$ of length $n$ is a sequence of vertices $v_0\cdots v_n$ such that $u=v_0$ and $w=v_n$, and such that $v_i\dof v_{i+i}$ for all $i\in\intset1{n-1}$.
A \defn{cycle} in $\gG$ is a path of positive length from a vertex $v$ to itself.
A graph $\gG$ is a \defn{directed acyclic graph} (or \defn{\dag}) if it contains no cycles.
A \defn{tree} is a \dag with a single vertex $r$ (called its \defn{root}) such that there is a path from any other vertex of the graph to $r$. A \defn{leaf} is a vertex which is not the target of any edge.

\begin{notation}
    We depict the edges of a graph $\gG$ by arrows between them.
    For example, the graph $\gG=\tuple{\set{a,b,c,d},\set{ab,bd,cd}}$ can be represented as
    $
    \va1\qquad\vb1\qquad\vc1\qquad\vd1
    \Dedges{a1/b1,c1/d1}
    \bentlDedges{b1/d1/20}
    $
    .
    In some cases, we assume the vertices have label naturally defined, in which case we represent them by their label.
\end{notation}

\section{Proof Nets}\label{sec:pn}

In this section we reformulate Lamarche's syntax for \emph{essential nets} \cite{lamarche:essential} for $\IMLL$ with units, and we show how to use them to define a canonical way to assign jumps to negative units, which yields a canonical representation of proofs in $\IMLL$.
\footnote{
    \defn{Essential nets} were defined for the unit-free fragment of propositional linear logic, but they already use jumps to represent the dependencies of the weakening rule. Removing exponentials and additives dramatically helps to simplify the presentation of essential nets, since it removes the need of slices for additives, and the way to recover the structure of boxes using the paths in the graph, which overload the syntax with additional definitions which are not required for the pure multiplicative case.
}

\begin{definition}
    A \defn{link} on a sequent $\Gamma\vdash A$ is a pair of nodes $\link=\tuple{v,w}$ such that $v$ is a negative leaf.
    It is an \defn{axiom link} if $v$ and $w$ are occurrences of the same atom, and a \defn{jump} if $v$ is an occurrence of $\lone$.

    A \defn{linking} $\links$ on a sequent $\Gamma\vdash A$ is a function associating a link to each negative leaf of $\Gamma\vdash A$.
    It is axiomatic if the restriction of $\links$ to the negative atoms forms a bijection between non-unit leaves.
\end{definition}

\begin{notation}
    We represent links by edges connecting the corresponding nodes in the sequent.
    In particular, we use squared edges to represent axiom links, and curved edges to represent jumps, as shown in the following example.
\end{notation}

\begin{example}
    Consider the sequent containing the formula \Cref{eq:sequent} in the consequent, and no formula in the antecedent, and the following linking $\links$ on it.
    \begin{equation}\label{eq:exampleLinking}
        \begin{array}{c}
        \\
            \Big(\big(((\vnp1 \ltens \vnone1)\limp \vp2)\limp (\vnq1\ltens\vnone3)\big)\ltens( \vlone4\limp \vnone5) \Big)\limp ((\vnp3\limp \vp4 )\vpz6{\ltens} \vq2)
            \sqDBedges{np1/p2/10,nq1/q2/12,np3/p4/10}
            \sqdDBedges{
                none1/p2/16,
                none3/pz6/16,
                none5/pz6/16
            }
        \\
        \end{array}
    \end{equation}
    
    In order to write it down as a set of pairs of nodes, we should annotate atoms and units to distinguish their occurrences (i.e., distinguish their nodes).
    For example, if we annotate the atoms, some units and connectives as follows
    $$
    \Big(\big(((p_1\ltens\lone_1)\limp p_2)\limp (q_1\ltens\lone_3)\big)\ltens( \lone\limp \lone_5) \Big)\limp ((p_3\limp p_4 ) {\ltens}_0 q_2)
    $$
    then $\links$ can be written as the following set of pairs of nodes
    $$\links=\set{\tuple{p_1,p_2},\tuple{p_3,p_4},\tuple{q_1,q_2},\tuple{\lone_1,p_2},\tuple{\lone_3,{\ltens}_0},\tuple{\lone_5,{\ltens}_0}}
    $$
\end{example}

\begin{figure}
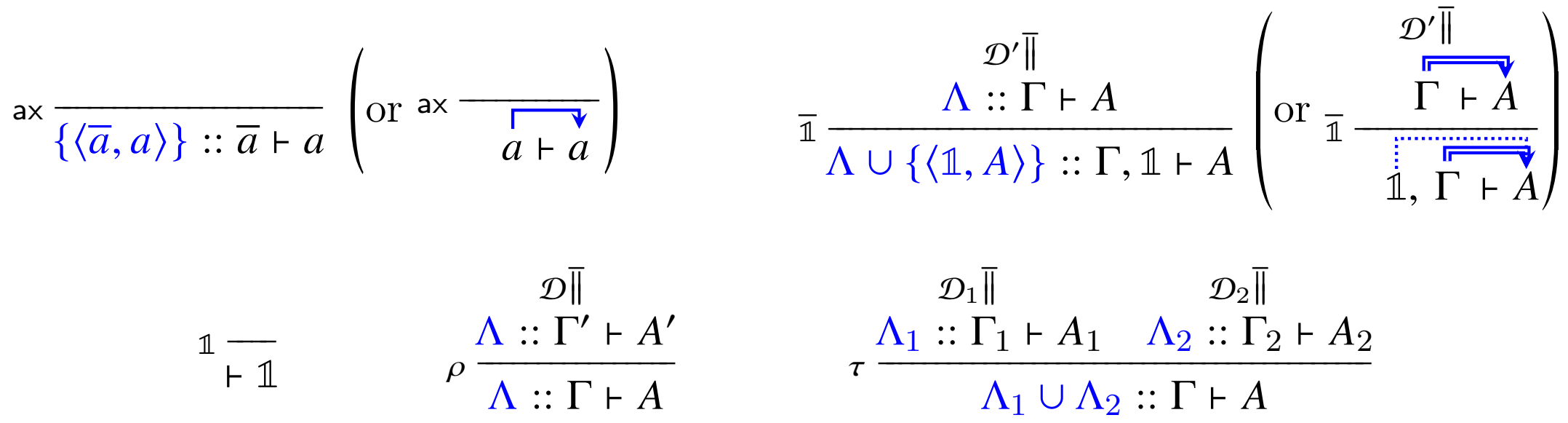

    \centering
    $$
    \begin{array}{c}
        \vlinf{\axrule}{}{\prenet{\set{\tuple{\cneg a, a}}}{\cneg a \vdash a}}{}
        \;
        \left(\text{or } 
            \vlinf{\axrule}{}{\phantom{\raisebox{5pt}{G}} \;\va1 \vdash \va2 \; \sqDBedges{a1/a2/10}}{}
        \right)
    \qquad
    \qquad
        \vlderivation{
                \vlin{\none}{}{\prenet{\links \cup \set{\tuple{\lone,A}}}{\Gamma,\lone \vdash A}}{
                    \vlpr{\dD'}{}{\prenet{\links}{\Gamma\vdash A}}
                }
            }
        \;
        \left(\text{or }
            \vlderivation{
            \vlin{\none}{}{
                \phantom{\raisebox{8pt}{G}} \vlone1, \vpz1{\Gamma}\vdash \vA1 \sqDBBedges{pz1/A1/10}
                \sqdDBedges{lone1/A1/13}
            }{
                \vlpr{\dD'}{}{\phantom{\raisebox{8pt}{G}}\vpz1{\Gamma}\vdash \vA1 \sqDBBedges{pz1/A1/10}}
            }
        }
        \right)
    \\\\
        \vlinf{\lone}{}{\vdash \lone}{}
    \qquad\qquad
        \vlderivation{
            \vlin{\rho}{}{\prenet{\links}{\Gamma \vdash A}}{
                \vlpr{\dD}{}{\prenet{\links}{\Gamma'\vdash A'}}
            }
        }
    \qquad\qquad
        \vlderivation{
            \vliin{\tau}{}{\prenet{\links[1]\cup\links[2]}{\Gamma\vdash A}}{
                \vlpr{\dD_1}{}{\prenet{\links[1]}{\Gamma_1\vdash A_1}}
            }{
                \vlpr{\dD_2}{}{\prenet{\links[2]}{\Gamma_2\vdash A_2}}
            }
        }
    \end{array}
    $$
    \caption{
        Translation of $\IMLL$ derivations into proof nets, with $\rho\in\set{\limp,\ntens}$ and $\tau\in\set{\nimp,\ltens}$.
        Note that linkings are defined over the nodes of the sequents, which are disjoint for disjoint sequents, thus the union is always disjoint.
    }
    \label{fig:pnTrans}
\end{figure}

We define proof nets as linkings obtained by translating derivations in $\IMLL$ or, equivalently, as linkings constructed by the rules in \Cref{fig:pnTrans}, mimicking the rules of the sequent calculus for $\IMLL$.

\begin{definition}
    A \defn{pre-net} is a pair $\genpren=\tuple{\Gamma\vdash A,\links}$ made of a sequent $\Gamma\vdash A$ and a linking $\links$ on $\Gamma\vdash A$.

    A \defn{proof net} for a sequent $\Gamma\vdash A$ is the pre-net $\genpren$ with linking $\links$ induced by a derivation $\dD$ in $\IMLL$  with conclusion $\Gamma\vdash A$ as follows:
    \begin{itemize}
        \item each $\axrule$ with conclusion $\cneg a\vdash a$ induces an axiomatic link $ \cneg a\vdash a$ in $\lambda$;
        \item each $\none$-rule with conclusion $\none, \Gamma \vdash A$ induces a jump link $\none \vdash A$ in $\lambda$.
    \end{itemize}
    Equivalently, a proof net is a pre-net generated by rules in \Cref{fig:pnTrans}.
\end{definition}

\begin{example}\label{ex:pnTrans}
    Below, a derivation in $\IMLL$ with the same conclusion is shown, together with its translation into a proof net, where we connect with an edge each pair of nodes in a same link (see \Cref{eq:testLink}).    

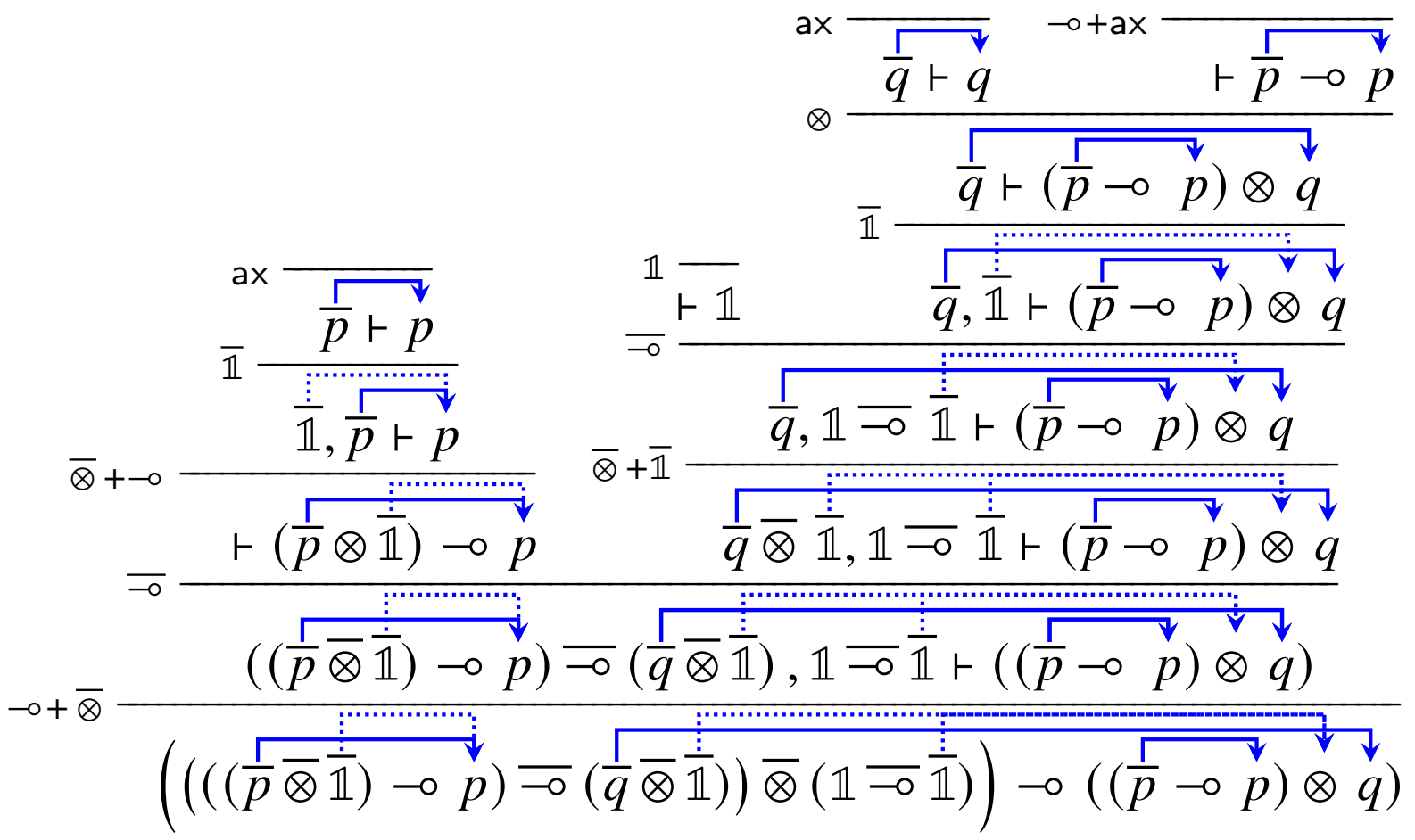
    \begin{equation}
        \vlderivation{
            \vlin{\limp+\ntens}{}{
                \phantom{\raisebox{10pt}{G}} 
                \Big(\big(((\vnp1 \ntens \vnone1) \limp \vp2) \nimp \, (\vnq1 \ntens\vnone3)\big)\ntens \, ( \vlone4\nimp \vnone5) \Big)\limp ((\vnp3 \limp \vp4 )\vpz9{\ltens} \vq2)
                \sqDBedges{np1/p2/10,nq1/q2/12,np3/p4/10}
                \sqdDBedges{none1/p2/14,none3/pz9/14,none5/pz9/14}
            }{
                \vliin{\nimp}{}{
                    \phantom{\raisebox{10pt}{G}} 
                    ((\vnp1 \ntens \vnone1) \limp \vp2) \nimp \, (\vnq1 \ntens \vnone3)
                    \, , \vlone4 \nimp  \vnone5
                    \vdash
                    ((\vnp3\vpz8{\limp} \vp4 )\vpz9{\ltens} \vq2)
                    \sqDBedges{np1/p2/10,nq1/q2/12,np3/p4/10}
                    \sqDBedges{nq1/q2/12,np3/p4/10}
                    \sqdDBedges{none5/pz9/14,none3/pz9/14,none1/p2/14}
                }{
                    \vlin{\ntens+\limp}{}{
                        \phantom{\raisebox{8pt}{G}} 
                        \vdash(\vnp1 \ltens \vnone1)\limp \vp2
                        \sqDBedges{np1/p2/10}
                        \sqdDBedges{none1/p2/12}
                    }{
                        \vlin{\none}{}{
                            \phantom{\raisebox{8pt}{G}} 
                            \vnone1,\vnp1 \vdash \vp2
                            \sqDBedges{np1/p2/10}
                            \sqdDBedges{none1/p2/12}
                        }{
                            \vlin{\axrule}{}{
                                \phantom{\raisebox{5pt}{G}} 
                                \vnp1 \vdash \vp2
                                \sqDBedges{np1/p2/10}
                            }{\vlhy{}}
                        }
                    }
                }{
                    \vlin{\ntens+\none}{}{
                        \phantom{\raisebox{10pt}{G}} 
                        \vnq1\vpz4{\ntens}\vnone3
                        , \vlone4\vpz6{\nimp} \vnone5
                        \vdash
                        (\vnp3\vpz8{\limp} \vp4 )\vpz9{\ltens} \vq2
                        \sqDBedges{nq1/q2/12,np3/p4/10}
                        \sqdDBedges{none5/pz9/14,none3/pz9/14}
                    }{
                        \vliin{\nimp}{}{
                            \phantom{\raisebox{10pt}{G}} 
                            \vnq1
                            , \vlone4\vpz6{\nimp} \vnone5
                            \vdash
                            (\vnp3\vpz8{\limp} \vp4 )\vpz9{\ltens} \vq2
                            \sqDBedges{nq1/q2/12,np3/p4/10}
                            \sqdDBedges{none5/pz9/14}
                        }{\vlin{\lone}{}{\vdash \lone}{\vlhy{}}}{
                            \vlin{\none}{}{
                                \phantom{\raisebox{10pt}{G}} 
                                \vnq1 , \vnone5
                                \vdash
                                (\vnp3\vpz8{\limp} \vp4 )\vpz9{\ltens} \vq2
                                \sqDBedges{nq1/q2/12,np3/p4/10}
                                \sqdDBedges{none5/pz9/14}
                            }{
                                \vliin{\ltens}{}{
                                    \phantom{\raisebox{8pt}{G}} 
                                    \vnq1
                                    \vdash
                                    (\vnp3\vpz8{\limp} \vp4 )\vpz9{\ltens} \vq2
                                    \sqDBedges{nq1/q2/12,np3/p4/10}
                                }{
                                    \vlin{\axrule}{}{
                                        \phantom{\raisebox{5pt}{G}} 
                                        \vnq1\vdash\vq2
                                        \sqDBedges{nq1/q2/10}
                                    }{}
                                }{
                                    \vlin{\limp+\axrule}{}{
                                        \phantom{\raisebox{5pt}{G}} 
                                        \vdash \vnp3\limp \vp4
                                        \sqDBedges{np3/p4/10}
                                    }{}
                                }
                            }
                        }
                    }
                }
            }
        }
    \end{equation}
\end{example}

We end this section by recalling the correctness criterion for Lamarche's proof nets for full intuitionistic linear logic in the case of $\IMLL$.
We slightly change the original definition of \test in \cite{lamarche:essential} by inverting the direction of the edges to ensure that linkings can always be considered as functions.
In fact, using the opposite direction a same vertex may be the source of more than one edge, as in the case of the (unique) proof net of the sequent $\lone,\lone\vdash\lone$.
\begin{definition}\label{def:correct}
    The \defn{\test} of a pre-net $\genpren$ is the directed graph $\testof\genpren$ with vertices the nodes of $\Gamma\vdash A$, and 
    edges defined either by the links in $\links$ (marked as $\vpz1{}\quad\vpz2{}\DBedges{pz1/pz2}$) as follows:
    \begin{equation}\label{eq:testLink}
        \hfill
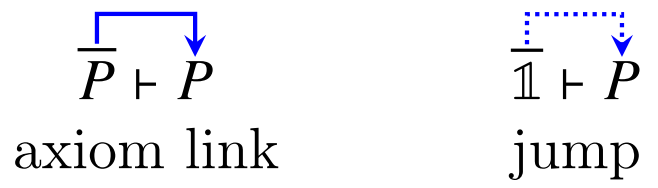
        \begin{array}{c@{\qquad\qquad}c}
        \\
            \vnP1 \vdash \vP1
            \sqDBedges{nP1/P1/12}
        &
            \vnone1 \vdash \vP1
            \sqdDBedges{none1/P1/12}
        \\
            \text{axiom link}
        &
            \text{jump}
        \end{array}
        \hfill
    \end{equation}
    or by the sub-formula relations (marked as $\vpz1{}\quad\vpz2{}\Aedges{pz1/pz2}$) as follows:
    \begin{equation}\label{eq:testForm}
        \hfill
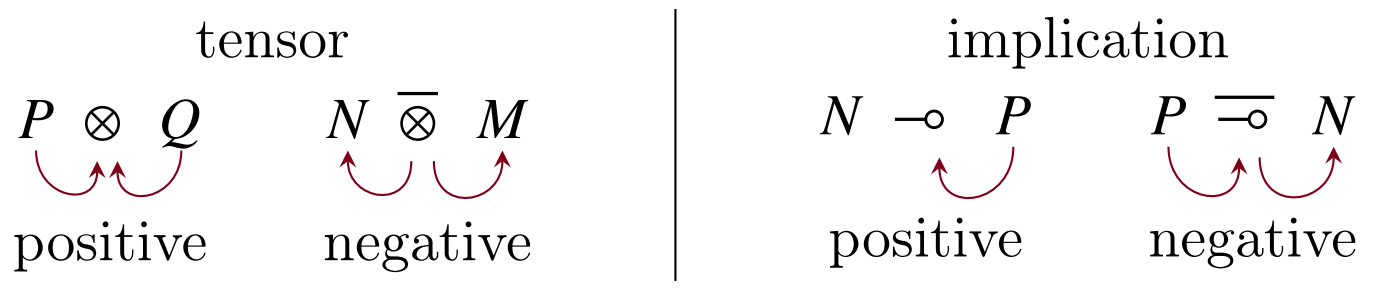
        \begin{array}{c@{\qquad\vrule\qquad}c}
            \text{tensor}
        &
            \text{implication}
        \\
            \begin{array}{c@{\qquad}c}
                \vP1 \;\vpz1{\ltens}\; \vQ1 
                \looseAedges{P1/pz1.-105/2/-90/-90,Q1/pz1.-75/2/-90/-90}
            &
                \vN1 \;\vpz1{\ntens}\; \vM1 
                \looseAedges{pz1.-105/N1/2/-90/-90,pz1.-75/M1/2/-90/-90}
            \\[10pt]
                \text{positive}
            &
                \text{negative}
            \end{array}
        &
            \begin{array}{c@{\qquad}c}
                \vN1 \;\vpz1{\limp}\; \vP1 
                \looseAedges{P1/pz1.-75/2/-90/-90}
            &
                \vP1 \;\vpz1{\nimp}\; \vN1
                \looseAedges{P1/pz1.-105/2/-90/-90,pz1.-75/N1/2/-90/-90}
            \\[10pt]
                \text{positive}
            &
                \text{negative}
            \end{array}
        \end{array}
        \hfill
    \end{equation}
    where source and target of the edges are the root nodes of the corresponding sub-formulas.
\end{definition}

\begin{remark}\label{rem:blackPaths}
    If $\Gamma\vdash A$ is a sequent, and $N$ is a negative sub-formula of $\Gamma\vdash A$, then there are always paths from $N$ to any negative leaf of the sub-tree of the parse tree of $N$.
\end{remark}

The correctness criterion for proof nets is defined in terms of the paths in the \test of a pre-net, as follows.
\begin{definition}
    Let $\genpren$ be a pre-net.
    It is \defn{correct} if it satisfies the following conditions:
    \begin{enumerate}
        \item\label{cor:acy} its \test is acyclic.
        \item\label{cor:conn} $B\pathto A$ for any sub-formula $B$ of $\Gamma\vdash A$.
        \item\label{cor:fun}
        if $L\limp R$ is a (positive) sub-formula in $\Gamma\vdash A$, then any maximal path in the \test starting from $L$ contains $R$.
        \footnote{
            By Condition \ref{cor:conn}, any such maximal path in a correct \test must end in the root of the consequent.
        }
    \end{enumerate}
    where by $B\pathto C$ we denote the existence of a path from (the root node of) $B$ to (the root node of) $C$ in the \test $\testof\genpren$.
\end{definition}

\begin{example}
    The \test of the linking in \Cref{eq:exampleLinking} is the following graph.
    
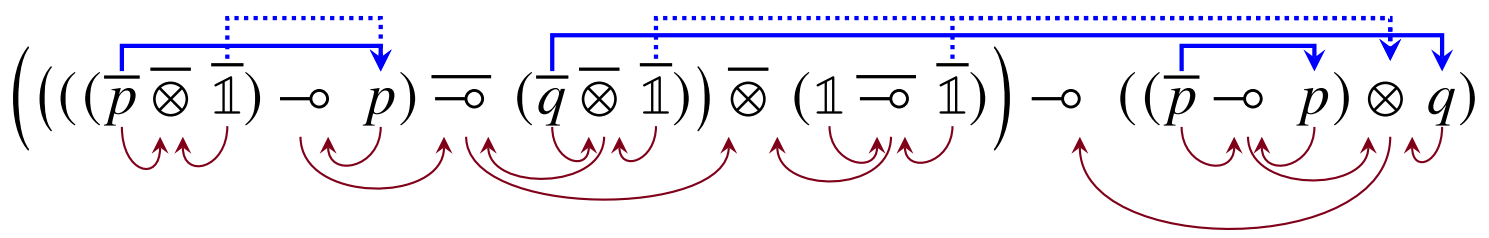
\begin{equation}\label{eq:exampleTest}
    \begin{array}{c}
        \\
        \Big(\big(((\vnp1 \vpz1{\ntens} \vnone1)\vpz2{\limp} \vp2)\vpz3{\nimp} (\vnq1\vpz4{\ntens}\vnone3)\big)\vpz5{\ntens}( \vlone4\vpz6{\nimp} \vnone5) \Big)\vpz7{\limp} ((\vnp3\vpz8{\limp} \vp4 )\vpz9{\ltens} \vq2)
        \\\\
    \end{array}
        \sqDBedges{np1/p2/10,nq1/q2/12,np3/p4/10}
        \sqdDBedges{none1/p2/14,none3/pz9/14,none5/pz9/14}
        \looseAedges{
            np1/pz1.-110/3/-90/-90,
            none1/pz1.-80/2.4/-90/-90,
            p2/pz2.-70/2/-90/-90,
            pz2.-110/pz3.-120/1.2/-90/-90,
            nq1/pz4.-110/2.4/-90/-90,
            none3/pz4.-70/2.4/-90/-90,
            pz4/pz3.-60/1.2/-90/-90,
            lone4/pz6.-110/2/-90/-90,
            none5/pz6.-70/2/-90/-90,
            pz6/pz5.-60/1.3/-90/-90,
            pz3/pz5.-120/.8/-90/-90,
            pz9/pz7.-70/1/-90/-90,
            np3/pz8.-110/2/-90/-90,
            p4/pz8.-70/2/-90/-90,
            pz8/pz9.-120/1.2/-90/-90,
            q2/pz9.-60/3/-90/-90
        }
    \end{equation}
\end{example}

\begin{remark}
    The edges defined by the sub-formula relation in \Cref{eq:testForm} can be seen as directing the connections in the parse tree of the sequent downward \resp{upward} if touching two positive nodes \resp{at least one negative node}, and removing the edge from the left child of a positive implication to its parent.
    
    For example, the \test in \Cref{eq:parse-tree} can be seen as the directed graph shown below, where the dotted edges are shown for uniformity (to keep track of the sub-formula relation), but they are not part of the \test.
    \begin{equation}\label{eq:polarizedParseTree}
    \begin{array}{ccc@{\quad}c@{\;}c@{\;}cccc@{\;}c@{\;}ccccccccccc}
        \\
            \vnp1&& \vnone1 & \vp4 && \vnq1 && \vnone3 && \vlone4 && \vnone5 & \vnp3 && \vp2 &&& \vq2
        \\
            &\vpz1{\ntens}
            &&&&
            &\vpz2{\ntens}
            &&&
            &\vpz3{\nimp}
            &&&
            \vpz4{\limp}
        \\
            &&
            \vpz5{\limp}
            &&&&&&
            &&&&&&&
            \vpz6{\ltens}
        \\
            &&&&\vpz7{\nimp}
        \\
            &&&&&&&\vpz9{\ntens}
        \\
            &&&&&&&&&&&\vpz8{\limp}
        \end{array}
        \Aedges{np1/pz1,none1/pz1,nq1/pz2,p4/pz5,none3/pz2,pz5/pz7,pz2/pz7,pz9/pz3}
        \Aedges{lone4/pz3,pz3/none5,p2/pz4,np3/pz4,q2/pz6,pz6/pz8,pz9/pz7,pz4/pz6}   
        \dashUAedges{pz5/pz1,pz8/pz9}
        \sqDBedges{np1/p4/10,nq1/q2/14,np3/p2/10}
        \sqdDBedges{none1/p4/12,none3/pz6/16,none5/pz6/16}
    \end{equation}
\end{remark}

\begin{theorem}\label{thm:correct}
    A pre-net $\genpren$ is a proof net if and only if it is correct.
\end{theorem}
\begin{proof}[Sketch of proof]
    We recall the main ideas of the proof from of the correctness criterion for essential nets from \cite{lamarche:essential}, which can be reformulated straightforwardly in our setting.

    The right-to-left direction is proved by induction on the structure of a derivation $\dD$ of the sequent $\Gamma\vdash A$, and by checking that the linking $\links[\dD]$ defined by the translation of $\dD$ is correct.
    
    The left-to-right direction is proved by showing that if $\genpren$ is a correct pre-net, then we can \emph{sequentialize} it, that is, we can construct a derivation $\dD$ of $\Gamma\vdash A$ such that $\links[\dD]=\links$.
    \begin{itemize}
        \item 
        If $\Gamma\vdash A$ is of the form $\vdash \lone$ or $a\vdash a$, then we can easily construct such derivation, which would be made of a single rule application of $\lone$ or $\axrule$ respectively.
    
        \item 
        Similarly, if $\Gamma$ contains a unit $\none$ or a $\ntens$-formulas, or if $A$ is a $\limp$-formula, then we can apply the corresponding rule to construct a derivation $\dD$ of $\Gamma\vdash A$ such that $\links[\dD]=\links$, and we can apply the induction hypothesis to the resulting sub-pre-net. Notice that the correctness condition \ref{cor:fun} ensures that if $A$ is a $\limp$-formula, then removing the main connective of $A$ will preserve the existence of paths from any sub-formula to the root of the consequent, which is required to apply the induction hypothesis.

        \item 
        If $\Gamma\vdash A$ is not of the above forms, then either $A=A_1\ltens A_2$ or a formula $A_1\nimp A_2$ in $\Gamma$ is \emph{splitting}, that is, such that removing the main connective of the formula the \test of the pre-net becomes disconnected into two components, each satisfying the correctness criterion. In this case, we apply the corresponding two-premises rule to the main connective of the splitting formula, and conclude by applying the induction hypothesis to the resulting sub-pre-nets.
    \end{itemize}

    To prove the existence of a splitting formula, it suffices to observe that the paths in the \test form a partial ordering on the sub-formulas of $\Gamma\vdash A$ (by conditions \ref{cor:conn} and \ref{cor:fun}), and that if none of the formulas in $\Gamma\vdash A$ is splitting, then there should be either a cycle, contradicting the correctness condition \ref{cor:acy}. 
\end{proof}

\section{Canonicity}\label{sec:canon}

In this section, we first recall the definition of proof equivalence in $\IMLL$ induced by rule permutations, showing that the proof nets defined in the previous section are not canonical with respect to this equivalence.
Then, we introduce the notion of \emph{maximality} for a linking, allowing us to define a canonical way to assign jumps, which yields a canonical representation of proofs in $\IMLL$.
For our canonical proof nets, we show a proof translation from sequent calculus derivations, and we explain how to define cut-elimination steps on them, which are strongly normalizing and confluent.

\subsection{Proof equivalence in $\IMLL$}\label{subsec:preq}

The standard way to define proof equivalence in a sequent calculus is by considering derivations to be equivalent modulo the permutations of \emph{independent} inference rules, where two inference rules are independent if they operate on different formulas of the sequents, that is, if their active and principal formulas are different and not in a sub-formula relation.

In \Cref{fig:rulesPermutation} we list the rule permutations defining the proof equivalence $\preq$ in $\IMLL\cup\set\cutr$ defined by the \emph{independent rule permutations}.
Note that for the unary-vs-binary right permutations the case $\ltens$-vs-$\limp$ is ruled out since both apply to the consequent of their sequent, and thus they cannot be independent.

\begin{figure}[t]
    \centering
    \adjustbox{max width= \textwidth}{$\begin{array}{r@{\quad\preq\quad}l}
        \multicolumn{2}{c}{\text{Unary-vs-unary}}
    \\[10pt]
        \vlderivation{
            \vlin{\mclr{\set{\none,\ntens}}}{}{
                \Gamma, \mclr{\Delta}, \Delta \vdash C
            }{
                \vlin{\aclr{\set{\limp,\none,\ntens}}}{}{
                    \Gamma, \mclr{\Delta}, \aclr{\Sigma} \vdash \aclr{C}
                }{
                    \vlhy{\Gamma, \Delta',\aclr{\Sigma'}\vdash \aclr{C'}}
                }
            }
        }
	&
	    \vlderivation{
            \vlin{\aclr{\set{\limp,\none,\ntens}}}{}{
                \Gamma, \Delta,\aclr{\Sigma}\vdash \aclr{C}
            }{
                \vlin{\mclr{\set{\none,\ntens}}}{}{
                    \Gamma, \mclr{\Delta}, \aclr{\Sigma'}\vdash \aclr{C'}
                }{
                    \vlhy{\Gamma, \mclr{\Delta'},{\Sigma'}\vdash C'}
                }
            }
        }
    \\[20pt]
        \multicolumn{2}{c}{\text{Binary left-vs-unary}}
    \\[10pt]
	    \vlderivation{
            \vlin{\aclr{\set{\none,\ntens}}}{}{
                \Gamma_1,\Gamma_2,\aclr{\Delta},\Sigma\vdash C
            }{
                \vliin{\mclr{\set{\nimp,\ltens,\cutr}}}{}{
                    \Gamma, \aclr{\Delta'}, \mclr\Sigma \vdash \mclr{C}
                }{
                    \vlhy{\Gamma_1, \Delta' \vdash \mclr{C_1}}
                }{
                    \vlhy{\Gamma_2,\mclr{\Sigma'}\vdash \mclr{C_2}}
                }
            }
        }
	&
	    \vlderivation{
            \vliin{\mclr{\set{\nimp,\ltens,\cutr}}}{}{
                \Gamma_1,\Gamma_2,{\Delta},\mclr{\Sigma}\vdash \mclr C
            }{
                \vlin{\aclr{\set{\none,\ntens}}}{}{
                    \Gamma_1, \aclr{\Delta} \vdash \mclr{C_1}
                }{
                    \vlhy{\Gamma_1, \aclr{\Delta'} \vdash \mclr{C_1}}
                }
            }{
                \vlhy{\Gamma_2,\mclr{\Sigma'}\vdash \mclr{C_2}}
            }
        }
    \\[20pt]
        \multicolumn{2}{c}{\text{Unary-vs-binary right}}
    \\[10pt]
       \vlderivation{
            \vliin{\mclr{\set{\ltens,\nimp,\cutr}}}{}{
                \Gamma,\mclr{\Delta},\Sigma\vdash \mclr C
            }{
                \vlhy{\Gamma_1, \vdash \mclr{C_1}}
            }{
                \vlin{\aclr{\set{\none,\ntens}}}{}{
                    \Gamma_2,\mclr{\Delta'},\aclr{\Sigma} \vdash \mclr{C_2}
                }{
                    \vlhy{\Gamma_2,{\Delta'},\aclr{\Sigma'} \vdash {C_2}}
                }
            }
        }
	&
	     \vlderivation{
            \vlin{\aclr{\set{\none,\ntens}}}{}{
                \Gamma,{\Delta},\aclr{\Sigma}\vdash C
            }{
                \vliin{\mclr{\set{\ltens,\nimp,\cutr}}}{}{
                    \Gamma,\mclr{\Delta},\aclr{\Sigma'} \vdash \mclr C
                }{
                    \vlhy{\Gamma_1 \vdash \mclr{C_1}}
                }{
                    \vlhy{\Gamma_2,\mclr{\Delta'},\Sigma' \vdash \mclr{C_2}}
                }
            }
        }
	\\[20pt]
        \multicolumn{2}{c}{\text{Binary left-vs-binary right}}
    \\[10pt]
	    \vlderivation{
            \vliin{\aclr{\rho_1}}{}{
                \Gamma_1,\Gamma_2,\Gamma_3,\aclr{\Delta},\Sigma \vdash C
            }{
                \vlhy{\Gamma_1\vdash \aclr{C_1}}
            }{
                \vliin{\mclr{\rho_2}}{}{
                    \Gamma_2,\Gamma_3, \aclr{\Delta'},\mclr\Sigma \vdash \mclr{C}
                }{
                    \vlhy{\Gamma_2,{\Delta'}\vdash \mclr{C_2}}
                }{
                    \vlhy{\Gamma_3,\mclr{\Sigma'}\vdash \mclr{C_3}}
                }
            }
        }
    &
        \vlderivation{
            \vliin{\mclr{\rho_2}}{}{
                \Gamma_1,\Gamma_2,\Gamma_3, \Delta, \mclr{\Sigma} \vdash \mclr C
            }{
                \vliin{\aclr{\rho_1}}{}{
                    \Gamma_1,\Gamma_2, \aclr{\Delta},\vdash \mclr{C_2}
                }{
                    \vlhy{\Gamma_1\vdash \aclr{C_1}}
                }{
                    \vlhy{\Gamma_2,\aclr{\Delta'}\vdash C_2}
                }
            }{
                \vlhy{\Gamma_3,\mclr{\Sigma'}\vdash \mclr{C_3}}
            }
        }
    \\[20pt]
        \vlderivation{
            \vliin{\aclr{\rho_1}}{}{
                \Gamma_1,\Gamma_2,\Gamma_3,\aclr{\Delta},\Sigma \vdash C
            }{
                \vlhy{\Gamma_1\vdash \aclr{C_1}}
            }{
                \vliin{\mclr{\rho_2}}{}{
                    \Gamma_2,\Gamma_3, \aclr{\Delta'},\mclr\Sigma \vdash \mclr{C}
                }{
                    \vlhy{\Gamma_2\vdash \mclr{C_2}}
                }{
                    \vlhy{\Gamma_3,{\Delta'},\mclr{\Sigma'}\vdash \mclr{C_3}}
                }
            }
        }
    &
        \vlderivation{
            \vliin{\mclr{\rho_2}}{}{
                \Gamma_1,\Gamma_2,\Gamma_3, \Delta, \mclr{\Sigma} \vdash \mclr C
            }{
                \vlhy{\Gamma_2\vdash \mclr{C_2}}
            }{
                \vliin{\aclr{\rho_1}}{}{
                    \Gamma_1,\Gamma_3, \aclr{\Delta},\mclr{\Sigma'}\vdash \mclr{C_3}
                }{
                    \vlhy{\Gamma_1\vdash \aclr{C_1}}
                }{
                    \vlhy{\Gamma_3,\aclr{\Delta'},{\Sigma'}\vdash {C_3}}
                }
            }
        }
    \end{array}$}
    \caption{
        Independent rule permutations in $\IMLL\cup\set\cutr$.
    }
    \label{fig:rulesPermutation}
\end{figure}

It is immediate to show that there are equivalent derivations in $\IMLL$ whose translations into proof nets are not the same.
In fact, the permutations binary left-vs-unary or unary-vs-binary right with the unary rule being a $\none$ can change the jump links, since they change the position of the jump associated to the unit introduced by the $\none$-rule (see \Cref{fig:pnTrans}).
For example, see \Cref{fig:introExample}.

\begin{proposition}\label{prop:semicanonical}
    Let $\dD_1$ and $\dD_2$ be two derivations in $\IMLL$.
    If $\dD_1\preq\dD_2$,
    the set of axiom links in $\links[\dD_1]$ and $\links[\dD_2]$ are the same, but the set of jump links in $\links[\dD_1]$ and $\links[\dD_2]$ may be different.
\end{proposition}

At the same time, it is not difficult to show that given two derivations $\dD_1$ and $\dD_2$ obtained by sequentializing the same proof net, then $\dD_1\preq\dD_2$.

\begin{lemma}\label{lem:sequentializationPreservesPreq}
    Let $\dD_1$ and $\dD_2$ be two derivations in $\IMLL$.
    If $\pnof{\dD_1}=\pnof{\dD_2}$, then $\dD_1\preq\dD_2$.
\end{lemma}

\subsection{Recovering Canonicity}\label{subsec:canon}
\def\extsto{\rightsquigarrow_{\mathsf{ext}}}
\def\extsfrom{\leftsquigarrow_{\mathsf{ext}}}
\def\extstos{\extsto^*}

To recover canonicity, we use the fact that the \test of a linking is a directed acyclic graph where every node $\nn$ admits a path to the root $\nr$ of the consequent to define a partial ordering of linkings.
Using this partial ordering, we can define a notion of \emph{extension} of a linking, and a notion of \emph{maximal linking} (i.e., a linking that cannot be extended).
By showing that the maximal extension of a linking is unique, and that the process of extending a linking is invariant with respect to the proof equivalence $\preq$, we can conclude that the maximal extension of the translation of a derivation is a canonical representation of the equivalence class of the derivation, and thus we obtain a way to assign jumps to negative units in a canonical way, which yields a canonical representation of proofs in $\IMLL$.

\begin{figure}
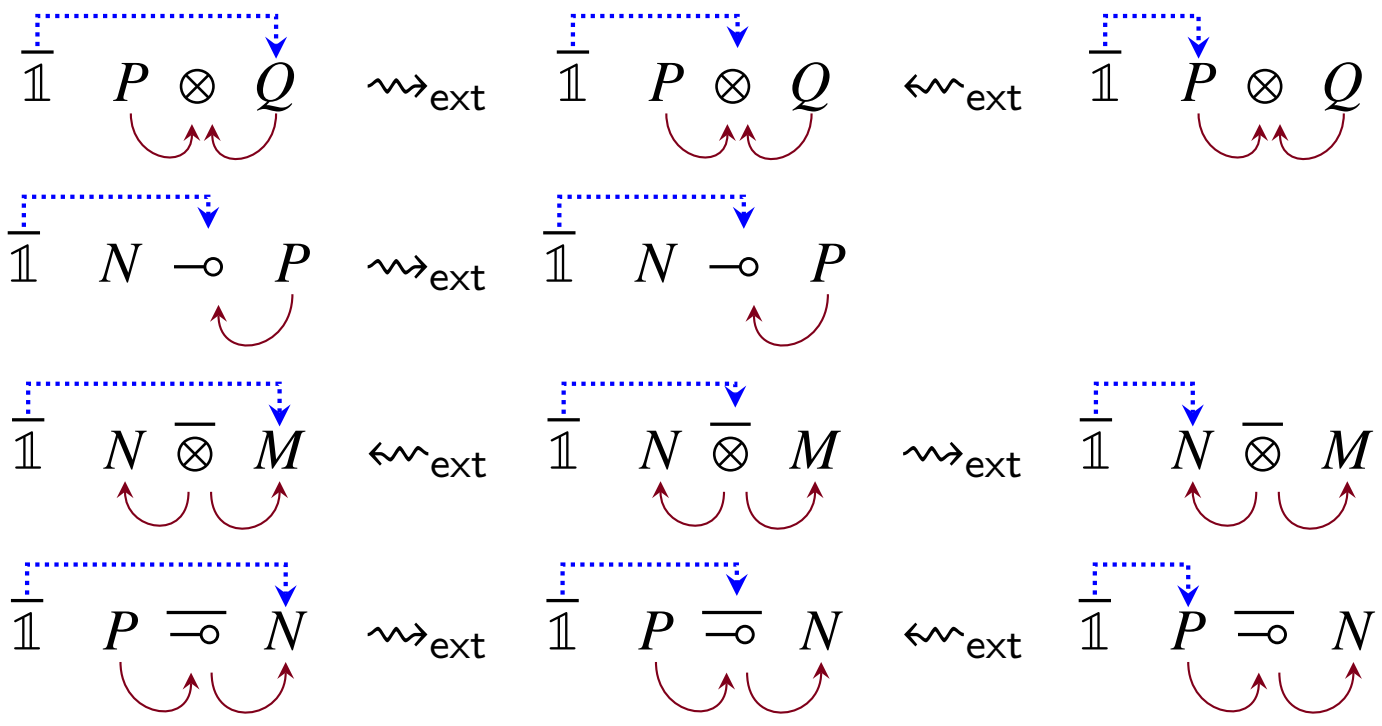

    \centering
    $$\begin{array}{ccccccc}
        \vnone1 \quad \vP1 \;\vpz1{\ltens}\; \vQ1 
        \looseAedges{P1/pz1.-105/2/-90/-90,Q1/pz1.-75/2/-90/-90}
        \sqdDBedges{none1/Q1/12}
    &\extsto&
        \vnone1 \quad \vP1 \;\vpz1{\ltens}\; \vQ1 
        \looseAedges{P1/pz1.-105/2/-90/-90,Q1/pz1.-75/2/-90/-90}
        \sqdDBedges{none1/pz1/12}
    &\extsfrom&
        \vnone1 \quad \vP1 \;\vpz1{\ltens}\; \vQ1 
        \looseAedges{P1/pz1.-105/2/-90/-90,Q1/pz1.-75/2/-90/-90}
        \sqdDBedges{none1/P1/12}
    \\[20pt]
        \vnone1 \quad \vN1 \;\vpz1{\limp}\; \vP1 
        \looseAedges{P1/pz1.-75/2/-90/-90}
        \sqdDBedges{none1/pz1/12}
    &\extsto&
        \vnone1 \quad \vN1 \;\vpz1{\limp}\; \vP1 
        \looseAedges{P1/pz1.-75/2/-90/-90}
        \sqdDBedges{none1/pz1/12}
    \\[20pt]
        \vnone1 \quad \vN1 \;\vpz1{\ntens}\; \vM1 
        \looseAedges{pz1.-105/N1/2/-90/-90,pz1.-75/M1/2/-90/-90}
        \sqdDBedges{none1/M1/12}
    &\extsfrom&
        \vnone1 \quad \vN1 \;\vpz1{\ntens}\; \vM1 
        \looseAedges{pz1.-105/N1/2/-90/-90,pz1.-75/M1/2/-90/-90}
        \sqdDBedges{none1/pz1/12}
    &\extsto&
        \vnone1 \quad \vN1 \;\vpz1{\ntens}\; \vM1 
        \looseAedges{pz1.-105/N1/2/-90/-90,pz1.-75/M1/2/-90/-90}
        \sqdDBedges{none1/N1/12}
    \\[20pt]
        \vnone1 \quad \vP1 \;\vpz1{\nimp}\; \vN1
        \looseAedges{P1/pz1.-105/2/-90/-90,pz1.-75/N1/2/-90/-90}
        \sqdDBedges{none1/N1/12}
    &\extsto&
        \vnone1 \quad \vP1 \;\vpz1{\nimp}\; \vN1
        \looseAedges{P1/pz1.-105/2/-90/-90,pz1.-75/N1/2/-90/-90}
        \sqdDBedges{none1/pz1/12}
    &\extsfrom&
        \vnone1 \quad \vP1 \;\vpz1{\nimp}\; \vN1
        \looseAedges{P1/pz1.-105/2/-90/-90,pz1.-75/N1/2/-90/-90}
        \sqdDBedges{none1/P1/12}
    \end{array}$$

    \caption{
        The local transformation of the jumps of a proof net when performing a \emph{simple extension}. The simple extensions on a $\ntens$ always induce a critical pair in the rewriting.
    }
    \label{fig:canonization}
\end{figure}

\begin{definition}\label{def:canon}
    Let $\links$ and $\links'$ be two linkings on the same sequent $\Gamma\vdash A$.
    We say that $\links'$ is a \defn{simple extension} of $\links$ if $\links'$ is obtained by shifting the target of a jump in $\links$ along an edge in the \test of $\links$, that is, if $\links'=(\links\setminus\set{\tuple{v,w}})\cup\set{\tuple{v,w'}}$ for some $\tuple{v,w}\in\links$ and $w'$ such that there is an edge $ww'$ in $\testof{\links}$.
    See \Cref{fig:canonization}.

    A linking $\links$ is \defn{maximal} if no simple extension of $\links$ is correct.
    A proof net $\genpren$ is \defn{canonical} if its linking is maximal.
\end{definition}

\begin{remark}\label{rem:rewiringJumps}
    If $\links$ is a linking on $\Gamma\vdash A$ and there is a jump in $\links$ with target a negative node $w$ of $\Gamma\vdash A$, then $\links$ is not maximal.
    This is because either 
    \begin{itemize}
        \item $w$ is a leaf of the parse tree of $\Gamma\vdash A$, in which case it must belong to an axiom link, and we can shift the jump to the other node of the axiom link;
        \item or $w$ is not a leaf of the parse tree of $\Gamma\vdash A$, in which case it must be either a $\ntens$-node or a $\nimp$-node, and we can always shift the jump to one of its children without violating the correctness Condition~\ref{cor:fun}.
    \end{itemize}
\end{remark}

\begin{example}
    The following proof net (where me marked the additional edges of its \test) is not maximal because the jump could be extended along the highlighted edge.
    $$\begin{array}{c}\\
        \left(\quad
        \vnone1 
        \quad\vpz1{\ntens}\quad
        (\quad\vlone 1 \quad\vpz2{\nimp}\quad \vna1\quad)
        \right)
        \quad \vpz3{\limp} \quad
        \va1
        \looseGOODedges{lone1/pz2.-110/1/-90/-90}
        \looseAedges{
            pz1.-110/none1/1.6/-90/-90,
            pz1.-70/pz2/.6/-90/-90,
            lone1/pz2.-110/1/-90/-90,
            pz2.-70/na1.-70/1/-90/-90,
            a1/pz3/1/-90/-90
        }
        \sqdDBedges{none1/lone1/12}
        \sqDBedges{na1/a1/12}
    \\[10pt]
    \end{array}$$
    since it can be extended by (repeatedly) shifting the jump as follows:
    $$
    \begin{array}{cc}
    \\
        \extsto
    &
        \left(\quad
        \vnone1 
        \quad\vpz1{\ntens}\quad
        (\quad\vlone 1 \quad\vpz2{\nimp}\quad \vna1\quad)
        \right)
        \quad \vpz3{\limp} \quad
        \va1
        \looseGOODedges{pz2.-70/na1.-70/1/-90/-90}
        \looseAedges{
            pz1.-110/none1/1.6/-90/-90,
            pz1.-70/pz2/.6/-90/-90,
            lone1/pz2.-110/1/-90/-90,
            pz2.-70/na1.-70/1/-90/-90,
            a1/pz3/1/-90/-90
        }
        \sqdDBedges{none1/pz2/12}
        \sqDBedges{na1/a1/12}
    \qquad\quad
    \\[25pt]
        \extsto
    &
        \left(\quad
        \vnone1 
        \quad\vpz1{\ntens}\quad
        (\quad\vlone 1 \quad\vpz2{\nimp}\quad \vna1\quad)
        \right)
        \quad \vpz3{\limp} \quad
        \va1
        \sqGOODedges{na1.60/a1/8}
        \looseAedges{
            pz1.-110/none1/1.6/-90/-90,
            pz1.-70/pz2/.6/-90/-90,
            lone1/pz2.-110/1/-90/-90,
            pz2.-70/na1.-70/1/-90/-90,
            a1/pz3/1/-90/-90
        }
        \sqdDBedges{none1/na1.120/12}
        \sqDBedges{na1.60/a1/8}
    \qquad\quad
    \\[25pt]
        \extsto
    &
        \left(\quad
        \vnone1 
        \quad\vpz1{\ntens}\quad
        (\quad\vlone 1 \quad\vpz2{\nimp}\quad \vna1\quad)
        \right)
        \quad \vpz3{\limp} \quad
        \va1
        \looseAedges{
            pz1.-110/none1/1.6/-90/-90,
            pz1.-70/pz2/.6/-90/-90,
            lone1/pz2.-110/1/-90/-90,
            pz2.-70/na1.-70/1/-90/-90,
            a1/pz3/1/-90/-90
        }
        \sqdDBedges{none1/a1/12}
        \sqDBedges{na1/a1/13}
    \qquad\quad
    \\[10pt]
    \end{array}
    $$

    Notice that the last linking is maximal, since a further extension would violate the correctness condition \ref{cor:fun}.
    In fact, the only possible extension of the last linking would be the following linking obtained by shifting the target of the jump to be the $\limp$-node, as shown below.
    $$\begin{array}{c}\\
        \left(\quad
        \vnone1 
        \quad\vpz1{\ntens}\quad
        (\quad\vlone 1 \quad\vpz2{\nimp}\quad \vna1\quad)
        \right)
        \quad \vpz3{\limp} \quad
        \va1
        \looseBADedges{pz1.-110/none1/1.6/-90/-90}
        \sqBADedges{none1/pz3/16}
        \looseAedges{
            pz1.-110/none1/1.6/-90/-90,
            pz1.-70/pz2/.6/-90/-90,
            lone1/pz2.-110/1/-90/-90,
            pz2.-70/na1.-70/1/-90/-90,
            a1/pz3/1/-90/-90
        }
        \sqdDBedges{none1/pz3/16}
        \sqDBedges{na1/a1/12}
    \\[10pt]
    \end{array}$$
    However, this would result in having a path in the \test from the $\ntens$-node (i.e., which is the root of the left argument of the $\limp$-node), to the root of the consequent (i.e., to the $\limp$-node itself) that does not contain the right argument of the $\limp$-node.
\end{example}

\begin{theorem}
    If $\links$ is a correct linking on $\Gamma\vdash A$, then there is a unique maximal linking which can be obtained from $\links$ by a sequence of simple extensions.
\end{theorem}
\begin{proof}
    We consider the relation $\extsto$ on linkings as a rewriting system, and we show that it is terminating and confluent.
    Termination follows immediately from the fact that the \test of $\links$ is a finite acyclic graph, and each simple extension shifts the target of a jump along an edge of the \test.

    Confluence is proven by showing that all critical pairs are joinable.
    If $\links[1]$ and $\links[2]$ are obtained by shifting two different jumps, then they are trivially joinable.
    If $\links[1]$ and $\links[2]$ are obtained by shifting the same jump $\tuple{v,w}$ to two different targets $w_1$ and $w_2$, then $w$ must have been a $\ntens$-node, and $w_1$ and $w_2$ must be the two children of $w$, both negative nodes.
    Let $u_1$ and $u_2$ be the targets of a jump obtained by shifting the jump $\tuple{v,w_1}$ and $\tuple{v,w_2}$ respectively, until possible, and let $\links[1]'$ and $\links[2]'$ be the resulting linkings.
    We know by \Cref{rem:rewiringJumps} that $u_1$ and $u_2$ must be positive nodes, and they cannot be children of $\ltens$-nodes, since we can always shift the jump without violating the correctness condition \ref{cor:fun}.
    If, wlog, $u_1\neq u_2$ is the positive root of $\Gamma\vdash A$ and $u_2$ is the right child of a $\limp$-node, then, since $\links[1]'$ is correct, we know that we should also be capable of safely shifting the jump $\tuple{v,u_2}$ since there is a path from the left child of such $\limp$-node to $v$ (passing through $w$) and following the sequence of vertices we used to extend the jump $\tuple{v,w}$ to $\tuple{v,u_1}$.
    With a similar reasoning, we can show that if $u_1$ is the right child of another $\limp$-node, then one between $\tuple{v,u_1}$ and $\tuple{v,u_2}$ can be safely shifted.
\end{proof}

Of course, one may find a canonical proof net for a sequent $\Gamma\vdash A$ by applying the translation in \Cref{fig:pnTrans} to any derivation of $\Gamma\vdash A$, and then by applying the simple extensions to the resulting proof net until we reach a maximal linking.
However, we can also define a direct translation from sequent calculus derivations to canonical proof nets as shown in \Cref{fig:canonTrans}.

\begin{figure}
    \centering

    $$
    \begin{array}{c}
        \vlinf{\axrule}{}{\;\va1 \vdash \va2 \; \sqDBedges{a1/a2/12}}{}
    \qquad
        \vlinf{\lone}{}{\vdash \lone}{}
    \qquad
        \vlderivation{
            \vlin{\rho}{}{\prenet{\max{(\links)}}{\Gamma \vdash A}}{
                \vlpr{\dD}{}{\prenet{\links}{\Gamma'\vdash A'}}
            }
        }
    \qquad
        \vlderivation{
            \vliin{\tau}{}{\prenet{\max(\links[1]\cup\links[2])}{\Gamma\vdash A}}{
                \vlpr{\dD_1}{}{\prenet{\links[1]}{\Gamma_1\vdash A_1}}
            }{
                \vlpr{\dD_2}{}{\prenet{\links[2]}{\Gamma_2\vdash A_2}}
            }
        }
    \end{array}
    $$

    \caption{
        Rules for the direct translation of $\IMLL$ derivations into proof nets replacing the corresponding rules in \Cref{fig:pnTrans}, where $\rho\in\set{\limp,\ntens}$ and $\tau\in\set{\nimp,\ltens}$.
        By $\max(\links)$ we denote the maximal linking obtained from $\links$ by applying simple extensions.
    }
    \label{fig:canonTrans}
\end{figure}

\begin{example}\label{ex:canonTrans}
    Below on the left, we show the the translation of a derivation into a canonical proof net using the translation in \Cref{fig:canonTrans}, and on the right we show the translation of the same derivation into a non-canonical proof net using the translation in \Cref{fig:pnTrans}.
    These are given by decorating each sequent in the derivation with the linking obtained by applying the translation to the derivation above the sequent.
    \begin{equation}
        \vlderivation{
            \vlin{\ntens}{}{
                \phantom{\raisebox{12pt}{G}} 
                \vnq1\vpz4{\ntens}\vnone3
                , \vlone4\vpz6{\nimp} \vnone5
                \vdash
                (\vnp3\vpz8{\limp} \vp4 )\vpz9{\ltens} \vq2
                \sqDBedges{nq1/q2/12,np3/p4/10}
                \sqdDBedges{none3/pz9/14,none5/pz9/14}
            }{
                \vliin{\nimp}{}{
                    \phantom{\raisebox{12pt}{G}} 
                    \vnq1,\vnone3
                    , \vlone4\vpz6{\nimp} \vnone5
                    \vdash
                    (\vnp3\vpz8{\limp} \vp4 )\vpz9{\ltens} \vq2
                    \sqDBedges{nq1/q2/12,np3/p4/10}
                    \sqdDBedges{none3/pz9/14,none5/pz9/14}
                }{
                    \vlin{\none}{}{
                        \phantom{\raisebox{10pt}{G}} 
                        \vnone3\vdash \vlone1
                        \sqdDBedges{none3/lone1/12}
                    }{\vlin{\lone}{}{\phantom{\raisebox{5pt}{G}} \vdash \vlone1}{\vlhy{}}}
                }{
                    \vliin{\ltens}{}{
                        \phantom{\raisebox{10pt}{G}} 
                        \vnq1 , \vnone5
                        \vdash
                        (\vnp3\vpz8{\limp} \vp4 )\vpz9{\ltens} \vq2
                        \sqDBedges{nq1/q2/12,np3/p4/10}
                        \sqdDBedges{none5/pz9/14}
                    }{
                        \vlin{\axrule}{}{
                            \phantom{\raisebox{5pt}{G}} 
                            \vnq1\vdash\vq2
                            \sqDBedges{nq1/q2/10}
                        }{}
                    }{
                        \vlin{\none}{}{
                            \phantom{\raisebox{5pt}{G}} 
                            \vnone5 \vdash \vnp3\limp\vp4
                            \sqDBedges{np3/p4/10}
                            \sqdDBedges{none5/p4/9}
                        }{
                            \vlin{\limp}{}{
                                \phantom{\raisebox{4pt}{G}} 
                                \vdash \vnp3\limp\vp4
                                \sqDBedges{np3/p4/10}
                            }{
                                \vlin{\axrule}{}{
                                    \phantom{\raisebox{4pt}{G}} \vnp3\vdash\vp4
                                    \sqDBedges{np3/p4/10}
                                }{}
                            }
                        }
                    }
                }
            }
        }
    \qquad
        \vlderivation{
            \vlin{\ntens}{}{
                \phantom{\raisebox{12pt}{G}} 
                \vnq1\vpz4{\ntens}\vnone3
                , \vlone1\vpz6{\nimp} \vnone5
                \vdash
                (\vnp3\vpz8{\limp} \vp4 )\vpz9{\ltens} \vq2
                \sqDBedges{nq1/q2/12,np3/p4/10}
                \sqdDBedges{none3/lone1/14,none5/p4/9}
            }{
                \vliin{\nimp}{}{
                    \phantom{\raisebox{12pt}{G}} 
                    \vnq1,\vnone3
                    , \vlone1\vpz6{\nimp} \vnone5
                    \vdash
                    (\vnp3\vpz8{\limp} \vp4 )\vpz9{\ltens} \vq2
                    \sqDBedges{nq1/q2/12,np3/p4/10}
                    \sqdDBedges{none3/lone1/14,none5/p4/9}
                }{
                    \vlin{\none}{}{
                        \phantom{\raisebox{10pt}{G}} 
                        \vnone3\vdash \vlone1
                        \sqdDBedges{none3/lone1/12}
                    }{\vlin{\lone}{}{\phantom{\raisebox{5pt}{G}} \vdash \vlone1}{\vlhy{}}}
                }{
                    \vliin{\ltens}{}{
                        \phantom{\raisebox{10pt}{G}} 
                        \vnq1 , \vnone5
                        \vdash
                        (\vnp3\vpz8{\limp} \vp4 )\vpz9{\ltens} \vq2
                        \sqDBedges{nq1/q2/12,np3/p4/10}
                        \sqdDBedges{none5/p4/9}
                    }{
                        \vlin{\axrule}{}{
                            \phantom{\raisebox{5pt}{G}} 
                            \vnq1\vdash\vq2
                            \sqDBedges{nq1/q2/10}
                        }{}
                    }{
                        \vlin{\none}{}{
                            \phantom{\raisebox{5pt}{G}} 
                            \vnone5 \vdash \vnp3\limp\vp4
                            \sqDBedges{np3/p4/10}
                            \sqdDBedges{none5/p4/9}
                        }{
                            \vlin{\limp}{}{
                                \phantom{\raisebox{4pt}{G}} 
                                \vdash \vnp3\limp\vp4
                                \sqDBedges{np3/p4/10}
                            }{
                                \vlin{\axrule}{}{
                                    \phantom{\raisebox{4pt}{G}} \vnp3\vdash\vp4
                                    \sqDBedges{np3/p4/10}
                                }{}
                            }
                        }
                    }
                }
            }
        }
    \end{equation}
    Notice that in the derivation on the left, 
    the top-most $\rltens$-rule and $\nimp$-rule shift the jumps to the $\ltens$-node, while in the derivation on the right the jumps are never shifted.
\end{example}

\begin{proposition}\label{prop:extensionPreservesPreq}
    Let $\genpren$ and $\genpren'$ be two proof nets for $\Gamma\vdash A$ such that $\links'$ is a simple extension of $\links$.
    Then $\genpren$ and $\genpren'$ are translations of equivalent derivations in $\IMLL$.
\end{proposition}
\begin{proof}
    It suffices to remark that each simple extension of a jump $\tuple{v,w}$ corresponds to the rule permutations in the sequent calculus moving the $\none$-rule introducing the node $v$ below the rule introducing the connective corresponding to the node $w$.
\end{proof}

We conclude the section by formally stating the main result of the paper.

\begin{theorem}[Canonicity]
    Let $\dD_1$ and $\dD_2$ be two derivations of $\Gamma\vdash A$ in $\IMLL$.
    Then $\dD_1\preq\dD_2$ iff $\cpnof{\dD_1}=\cpnof{\dD_2}$.
\end{theorem}
\begin{proof}
    If $\dD_1\preq\dD_2$, then by definition there is a sequence of independent rule permutations from \Cref{fig:rulesPermutation} transforming $\dD_1$ into $\dD_2$. 
    We conclude by induction on the number of these permutations, observing that rules not involving $\none$-rules do not change the linking, while the rule permutations involving $\none$-rules and moving a $\none$-rule below a rule changing the formula in the consequent (i.e., a $\ltens$, $\limp$, or a $\nimp$-rule) correspond to simple extensions of the linking (see \Cref{prop:semicanonical} and \Cref{prop:extensionPreservesPreq})

    To prove the converse, we show by induction on the number of vertices in the \test of $\genpren=\cpnof{\dD_1}=\cpnof{\dD_2}$ that if it is possible to sequentialize $\genpren$ into a derivation $\dD_1$ and into a derivation $\dD_2$ starting with two different rules, then there are two derivations $\dD_1'$ and $\dD_2'$ such that $\dD_1'\preq\dD_2'$, and such that $\dD_1'$ and $\dD_2'$ only differ in the order of the last two rules, which can be permuted by one of the rule permutations in \Cref{fig:rulesPermutation}.
\end{proof}

\subsection{On Cut Elimination for Canonical Proof Nets}\label{sec:canoncut}
\def\ccut{\overset{\cutr}{\limp}}

We conclude this section by briefly discussing how to define cut-elimination for canonical proof nets.
Since we are defining proof nets as decorations of sequents, we have to include a connective in the language of formulas to represent a pair of cut-formulas in the sequent, and to extend the definition of \test to include such connective.

The results in the previous section are straightforwardly extended to canonical proof nets with cuts, by simply considering the $\ccut$-connective as a $\nimp$, except for the definition of the simple extensions, where we forbid to shift the target of a jump to a cut-formula.

\begin{definition}
    The \defn{cut-connective} $\ccut$ is a binary connective.
    A \defn{cut-formula} is a formula of the form $A\ccut A$ where $A$ is a regular formula.
    A \defn{sequent with cuts} is a sequent $\Gamma\vdash A$ where $\Gamma$ may contain cut-formulas.
    The cut-connective should be considered as a $\nimp$ with respect to the definition of the polarity of formulas.

    A \defn{pre-net with cuts} $\genpren$ is given by a linkings $\links$ on a sequent with cuts.
    A \defn{simple extension} of a linking $\links$ on a sequent with cuts is defined as in \Cref{def:canon}, but forbidding to shift the target of a jump to the node corresponding to the cut-connective.
    A linking is \defn{cut-maximal} if it does not admit a simple extension that is correct, or that shift the target of a jump to a cut-formula.

    A \defn{canonical net with cuts} is a pre-net obtained by translating a derivation in $\IMLL$ with cuts using the translation in \Cref{fig:canonTrans} extended with the following rule for the cut rule:
    
    \begin{equation}
        \vlderivation{
            \vliin{\cutr}{}{
                \prenet{\links[1]\cup\links[2]}{A\ccut A,\Gamma,\Delta\vdash C}}{
                \vlpr{\dD_1}{}{\prenet{\links[1]}{\Gamma\vdash A}}}{
                \vlpr{\dD_2}{}{\prenet{\links[2]}{A,\Delta\vdash C}}
            }
        }
    \end{equation}

    A pre-net with cuts is a \defn{correct} if it satisfies the correctness criterion in \Cref{def:correct} with the following extension for the definition of the edges in \Cref{eq:testForm} for the cut-connetive nodes:
    \begin{equation}
        \hfill
        \begin{array}{c}
            \text{cut-connective}
        \\
            \vP1 \;\vpz1{\ccut}\; \vN1
        \\
            \looseAedges{P1/pz1.-105/2/-90/-90,pz1.-75/N1/2/-90/-90}
        \end{array}
        \hfill
    \end{equation}    
\end{definition}

We define the cut-elimination steps for canonical proof nets, which are the same as the cut-elimination steps for non-canonical proof nets, but with the additional condition that the resulting linking after performing the cut-elimination step is a simple extension of the original linking.
\begin{equation}
    \begin{array}{r@{\quad\cutelim\quad}l}
        \prenet{\links}{\Gamma, \lone\ccut\lone\vdash C} 
        &
        \prenet{\max{(\links)}}{\Gamma \vdash C}      
    \\
        \prenet{\links}{\Gamma, a\ccut a\vdash C} 
        &
        \prenet{\max{(\links)}}{\Gamma \vdash C}   
    \\
        \prenet{\links}{\Gamma, (\isb A\limp \isw B) \ccut (\isw A \nimp \isb B) \vdash C} 
        &
        \prenet{\max{(\links)}}{\Gamma, \isw A \ccut \isb A, \isw B \ccut \isb B \vdash C} 
    \\
        \prenet{\links}{\Gamma, (\isw A\ltens \isw B) \ccut (\isb A \ntens \isb B) \vdash C} 
        &
        \prenet{\max{(\links)}}{\Gamma, \isw A \ccut \isb A, \isw B \ccut \isb B \vdash C} 
    \end{array}
\end{equation}
Notice that the last two cut-elimination step only shift the target of those jumps that had the target the positive connective of the positive cut-formula.

\begin{proposition}
    Each canonical cut-elimination step on a proof net can be performed in quadratic time in the size of the proof net.
\end{proposition}
\begin{proof}
    The number of jumps in the proof net is linear in the size of the sequent, and they can be shifter a number of times which is bound by the number of nodes in the \test, which is also linear in the size of the sequent.
\end{proof}

\section{Jumpless Proof Nets}\label{sec:jumpless}

In this section, we show that, contrary to the classical case, where there cannot be a polynomial-time correctness criterion for proof nets without jumps characterizing the class of proof nets corresponding to correct proofs in $\MLL$, such result is possible in the intuitionistic case, and we provide a proof of it.
In particular, we show that it is possible to weaken the second and third condition of the correctness criterion in \Cref{def:correct} to obtain a correctness criterion that still characterizes proof nets for $\IMLL$, but which does not require considering jumps.

\begin{definition}
    A \defn{jumpless linking} $\links$ for a sequent $\Gamma\vdash A$ is a map from the negative non-unit leaves to the positive leaves of $\Gamma\vdash A$ inducing a bijection between negative and positive non-unit leaves.
    A \defn{jumpless net} $\genpren$ for a sequent $\Gamma\vdash A$ is a jumpless linking $\links$ on $\Gamma\vdash A$ which is the restriction of a linking defined as in \Cref{fig:pnTrans} to axiomatic links, i.e., removing jumps.
\end{definition}

\begin{definition}\label{def:corrJumpless}
    Let $\genpren$ be a pre-net for $\Gamma\vdash A$.
    It is \defn{weakly correct} if it satisfies the following conditions:
    \begin{enumerate}
        \item its \test is acyclic.
        \item\label{cor:wfun} if $L\limp R$ is a sub-formula in $\Gamma\vdash A$, then all maximal paths starting from $L$ ending in a positive node contains $R$.
    \end{enumerate}
\end{definition}
Intuitively, this weaker version of Condition~\ref{cor:fun} is due to the fact that, removing jumps, Condition~\ref{cor:conn} does not hold anymore. Then, each maximal path would then either in $A$ or in a $\none$. 
In particular, if all maximal paths will end in some $\none$'s, then we can take the greatest positive gate they reach to be intended as the right-hand side formula of the greatest derivable sequent containing the formula $L\limp R$ (see the notion of empire in \cite{bellin:subnets}).

To prove the fact that the weakly correct criterion characterizes jumpless nets, we need to show that the \emph{splitting lemma}, which we mention in the sketch of the  proof of \Cref{thm:correct}, holds for pre-nets satisfying the weaker correctness criterion.
\begin{lemma}[Splitting]\label{lem:splitting}
    Let $\genpren$ be a weakly correct pre-net for a sequent $\Gamma\vdash A$ not of the form $a\vdash a$ and such that $\Gamma$ does not contain $\none$ or a $\ntens$, and $A$ is not of the form $N\limp P$.
    Then, 
    \begin{enumerate}
        \item\label{split:imp} either there is a splitting formula $P\limp N$ in $\Gamma$, that is, 
        $\Gamma=\Gamma_1, P\nimp N, \Gamma_2$ and there are two disjoint sets of links $\links[1]$ and $\links[2]$ such that $\links=\links[1]\cup\links[2]$, and $\links[1]$ only contains sub-sequents of $\Gamma_1\vdash P$ and $\links[2]$ only contains sub-sequents of $\Gamma_2, N \vdash A$;

        \item\label{split:tens} or $A$ is splitting, that is, $A=P_1\ltens P_2$ and $\Gamma=\Gamma_1, \Gamma_2$, and there are two disjoint sets of links $\links[1]$ and $\links[2]$ such that $\links=\links[1]\cup\links[2]$, and $\links[i]$ only contains sub-sequents of $\Gamma_i\vdash P_i$ for $i\in\set{1,2}$.
    \end{enumerate}
\end{lemma}
\begin{proof}
    We consider the partial order over the subformulas in $\Gamma\vdash A$ defined by the $\pathto$-paths in the \test by letting $B<C$ if $C$ occurs in all $\pathto$-path from $B$ to $A$ in the \test.

    If a formulas in $\Gamma$ admits no $\pathto$-paths to $A$, then we can take the $<$-maximal formula among them, which must be a splitting.
    It cannot be a (negative) atom by definition of $\pathto$, and it cannot be $\none$ nor a $\ntens$ by hypothesis on $\Gamma$.
    Therefore, such maximal formula must be a $\limp$-formula $P\limp N$.
    We conclude as in Case~\ref{split:imp} by letting $\Gamma_1$ be made of those formulas in $\Gamma$ which are $<$-smaller than $P$.

    Otherwise, the argument is similar to the one from \cite{lamarche:essential}:
    pick any $\limp$-formula $P\limp N$ in $\Gamma$ and consider the set of formulas in $\Gamma$ which are smaller than $P\limp N$. If the maximal (with respect to the subformula order) formulas $<$-smaller than $P$ are all $\limp$-formulas, then we can let $\Gamma_1$ be made of those formulas in $\Gamma$ which are $<$-smaller than $P$, and we conclude as in Case~\ref{split:imp}.
    If there is a maximal subformula $<$-smaller than $P$ which is not a in $\Gamma$, then it must be a subformula of a $\limp$-formula $P'\limp N'$ in $\Gamma$, and we can repeat the argument on $P'$. 
    By acyclicity of the \test, this process must terminate.
    If we found a formula $P\limp N$ in $\Gamma$ such that all the maximal formulas $<$-smaller than $P$ are formulas in $\Gamma$, then we conclude as in Case~\ref{split:imp}.
    Otherwise, $\Gamma$ must contain atomic formulas only, and $A=P_1\ltens P_2$, and we conclude as in Case~\ref{split:tens} by letting $\Gamma_i$ be made of those formulas in $\Gamma$ which are $<$-smaller than $P_i$ for $i\in\set{1,2}$.
\end{proof}

\begin{theorem}\label{thm:jumplessCorrect}
    A pre-net is a jumpless net iff it is weakly correct.
\end{theorem}
\begin{proof}
    The structure of the proof is similar to the proof of \Cref{thm:correct}, relying on the fact that for the splitting lemma (\Cref{lem:splitting}) to hold it suffices that the pre-net satisfies the weakly-correctness criterion.
\end{proof}

\section{Conclusion}\label{sec:conc}

In this paper, we answer positively a longstanding open question about the theory of proof nets for intuitionistic linear logic:
\begin{quotation}
    \emph{
        Is it possible to define a canonical representation of proofs in $\IMLL$
    \\
        (with $\comPtime$ proof translation and $\comPtime$ correctness criterion)?
    }
\end{quotation}
To answer this question, we have defined a notion of canonical proof nets for $\IMLL$ by requiring that jumps satisfy local conditions, and we have shown that any proof net can be transformed into a canonical one by performing simple local rewriting steps with linear complexity.
To this aim, we relied on Lamarche's correctness criterion for proof nets for intuitionistic linear logic, which is a refinement of the Danos-Regnier switching criterion for classical multiplicative proof nets, and we show that the same directed paths used by the correctness criterion can be used to guide the rewriting steps to transform a general proof net into a canonical one.

\paragraph{Implications and Future Work}

As a consequence of our canonicity result, the same approach could be used to define canonical proof nets for the multiplicative exponential fragment of intuitionistic linear logic ($\IMELL$).
This also opens to the possibility of defining canonical proof nets for fragments of (classical) multiplicative exponential linear logic ($\MELL$) used to encode lambda calculus, addressing the current unsatisfactory situation where the syntax of proof nets for $\MELL$ requiring the mix to be taken into account.

Another curious consequence of \Cref{sec:jumpless} is that, differently from the classical case, in $\IMLL$ there is a polynomial-time correctness criterion for proof nets without jumps, which is conservative over the unit-free case.
In fact, for classical proof nets, a simple correctness criterion for proof nets with units but without jumps cannot exists, and isolate fragments of classical logic admitting such criterion  (see \cite{regnier:phd,did:tor:25}).
This results also ensures that the category of jump-free proof nets for $\IMLL$ (with object formulas and morphisms between $A$ and $B$ are (jump-free) proof nets for $A\limp B$) is a non-trivial quotient of the category of \smccs, which is conservative over morphisms between unit-free objects.
In such a quotient, differently from free \smccs, the following diagram from \cite{kel:mac:71} commutes.

\begin{equation}\label{eq:diagram}
    \hfill
    \begin{array}{c@{\hskip4em}c@{\hskip4em}c}
        \vpz1{((A\limp \lone)\limp \lone) \limp\lone} 
    &&
        \vpz3{((A\limp \lone)\limp \lone) \limp \lone}
    \\[30pt] 
    &
        \vpz2{A\limp \lone}
    \end{array}
    \tikz[overlay,remember picture,draw,fill,opacity=1]{
        \draw[>=stealth,->,draw, thick] (pz1) -- (pz2) node[midway,below] {$\mathsf k_{A\limp \lone}$};
        \draw[>=stealth,->,draw, thick] (pz1) -- (pz3) node[midway,above] {$\mathsf{id}$} ;
        \draw[>=stealth,->,draw, thick] (pz2) -- (pz3) node[midway,below] {\qquad$\mathsf k^{-1}_{A} \limp \lone$};
    }
    \hfill
\end{equation}
where $\mathsf k_{X}$ is the canonical isomorphism between from $(X\limp \lone)\limp \lone$ to $X$ in the free \smcc, and $\mathsf k^{-1}_{X}$ is its inverse.

\bibliographystyle{plain}
\bibliography{biblio}

@article{kel:mac:71,
  title = {Coherence in closed categories},
  journal = {Journal of Pure and Applied Algebra},
  volume = {1},
  number = {1},
  pages = {97-140},
  year = {1971},
  issn = {0022-4049},
  doi = {https://doi.org/10.1016/0022-4049(71)90013-2},
  url = {https://www.sciencedirect.com/science/article/pii/0022404971900132},
  author = {G.M. Kelly and S. Maclane}
}

@article{bellin:subnets,
	title={Subnets of proof-nets in multiplicative linear logic with MIX},
	author={Bellin, Gianluigi},
	journal={Mathematical Structures in Computer Science},
	volume={7},
	number={6},
	pages={663--669},
	year={1997},
	publisher={Cambridge University Press}
}

@incollection{acc:str:AIML22,
  author       = {Matteo Acclavio and Lutz Straßburger},
  title        = {Combinatorial Proofs for Constructive Modal Logic},
  booktitle    = {Advances in Modal Logic, Volume 14},
  editor       = {David Fernández-Duque, Alessandra Palmigiano and Sophie Pinchinat},
  pages        = {15--36},
  publisher    = {College Publications},
  year         = {2022},                            
  url          = {http://www.aiml.net/volumes/volume14/06-Acclavio-Strassburger.pdf}
}

@article{reconfiguration,
  title = {On the complexity of reconfiguration problems},
  journal = {Theoretical Computer Science},
  volume = {412},
  number = {12},
  pages = {1054-1065},
  year = {2011},
  issn = {0304-3975},
  doi = {https://doi.org/10.1016/j.tcs.2010.12.005},
  url = {https://www.sciencedirect.com/science/article/pii/S0304397510006961},
  author = {Takehiro Ito and Erik D. Demaine and Nicholas J.A. Harvey and Christos H. Papadimitriou and Martha Sideri and Ryuhei Uehara and Yushi Uno}
}

@INPROCEEDINGS{constraintLogic,
  author={Demaine, Erik D. and Hearn, Robert A.},
  booktitle={2008 23rd Annual IEEE Conference on Computational Complexity}, 
  title={Constraint Logic: A Uniform Framework for Modeling Computation as Games}, 
  year={2008},
  volume={},
  number={},
  pages={149-162},
  doi={10.1109/CCC.2008.35}}

@article{acc:EHPN, 
	doi = {10.4204/eptcs.353.1},
	url = {https://doi.org/10.4204%2Feptcs.353.1}, 
	year = 2021, 
	month = {dec}, 
	publisher = {Open Publishing Association}, 
	volume = {353}, 
	pages = {1--25}, 
	author = {Matteo Acclavio}, 
	title = {Exponentially Handsome Proof Nets and Their Normalization}, 
	journal = {Electronic Proceedings in Theoretical Computer Science}
}

@inproceedings{acc:str:IJCAR18,
  title={From Syntactic Proofs to Combinatorial Proofs},
  author={Acclavio, Matteo and Stra{\ss}burger, Lutz},
  booktitle={International Joint Conference on Automated Reasoning},
  pages={481--497},
  year={2018},
  organization={Springer}
}

@InProceedings{dig:fag:jumps,
  author="Di Giamberardino, Paolo
  and Faggian, Claudia",
  editor="{\'E}sik, Zolt{\'a}n",
  title="Jump from Parallel to Sequential Proofs: Multiplicatives",
  booktitle="Computer Science Logic",
  year="2006",
  publisher="Springer Berlin Heidelberg",
  address="Berlin, Heidelberg",
  pages="319--333",
  isbn="978-3-540-45459-5"
}

@phdthesis{regnier:phd,
  title={Lambda-calcul et r{\'e}seaux},
  author={Regnier, Laurent},
  year={1992},
  school={Paris 7}
}

@misc{did:tor:25,
  title={On the role of connectivity in Linear Logic proofs}, 
  author={Raffaele Di Donna and Lorenzo Tortora de Falco},
  year={2025},
  eprint={2506.21678},
  archivePrefix={arXiv},
  primaryClass={cs.LO},
  url={https://arxiv.org/abs/2506.21678}, 
}

@article{gir:ll,
  title={Linear logic},
  author={Girard, Jean-Yves},
  journal={Theoretical computer science},
  volume={50},
  number={1},
  pages={1--101},
  year={1987},
  publisher={Elsevier}
}

@incollection{girard:PN,
  title={Proof-nets: the parallel syntax for proof-theory},
  author={Girard, Jean-Yves},
  booktitle={Logic and Algebra},
  pages={97--124},
  year={2017},
  publisher={Routledge}
}

@article{hughes:pws,
  author    = {Dominic J.D. Hughes},
  title     = {Proofs {W}ithout {S}yntax},
  journal   = {Annals of Math.},
  year      = 2006,
  volume    = 164,
  number    = 3,
  pages     = "1065--1076"
}

@Article{danos:regnier:89,
  author = "Vincent Danos and Laurent Regnier",
  title = "The structure of multiplicatives",
  journal = aml,
  volume = 28,
  pages = "181--203",
  year = 1989,
}

@article{hughes:invar,
  author    = {Dominic J.D. Hughes},
  title     = {Towards {H}ilbert's 24\({}^{\mbox{th}}\) Problem: Combinatorial Proof
               Invariants: (Preliminary version)},
  journal   = {Electr. Notes Theor. Comput. Sci.},
  volume    = {165},
  pages     = {37--63},
  year      = {2006},
}

@misc{hughes:fopws,
  archivePrefix = {arXiv},
  eprinttype = {arxiv},
  eprint = {1906.11236},
  primaryClass = {math},
  title = {First-Order Proofs without Syntax},
  journal = {arXiv:1906.11236 [math]},
  author = {Hughes, Dominic J. D.},
  month = jun,
  year = {2019},
}

@InProceedings{acc:str:CPK,
	author="Acclavio, Matteo and Stra{\ss}burger, Lutz",
	editor="Cerrito, Serenella and Popescu, Andrei",
	title="On Combinatorial Proofs for Modal Logic",
	booktitle="Automated Reasoning with Analytic Tableaux and Related Methods",
	year="2019",
	publisher="Springer International Publishing",
	address="Cham",
	pages="223--240",
	isbn="978-3-030-29026-9"
}

@inproceedings{ICP,
  author={Heijltjes, Willem B. and Hughes, Dominic J. D. and StraBburger, Lutz},
  booktitle={2019 34th Annual ACM/IEEE Symposium on Logic in Computer Science (LICS)}, 
  title={Intuitionistic proofs without syntax}, 
  year={2019},
  volume={},
  number={},
  pages={1-13},
  doi={10.1109/LICS.2019.8785827}
}

@inproceedings{hei:hug:conflict,
author = {Hughes, Dominic and Heijltjes, Willem},
title = {Conflict nets: Efficient locally canonical MALL proof nets},
year = {2016},
isbn = {9781450343916},
publisher = {Association for Computing Machinery},
address = {New York, NY, USA},
url = {https://doi.org/10.1145/2933575.2934559},
doi = {10.1145/2933575.2934559},
booktitle = {Proceedings of the 31st Annual ACM/IEEE Symposium on Logic in Computer Science},
pages = {437-446},
numpages = {10},
location = {New York, NY, USA},
series = {LICS '16}
}

@inproceedings{heijltjes:houston:14,
	author    = {Willem Heijltjes and
	Robin Houston},
	title     = {No proof nets for {MLL} with units: proof equivalence in {MLL} is
	{PSPACE}-complete},
	editor    = {Thomas A. Henzinger and
	Dale Miller},
	booktitle = {Joint Meeting of the Twenty-Third {EACSL} Annual Conference on Computer
	Science Logic {(CSL)} and the Twenty-Ninth Annual {ACM/IEEE} Symposium
	on Logic in Computer Science (LICS), {CSL-LICS} '14, Vienna, Austria,
	July 14 - 18, 2014},
	pages     = {50:1--50:10},
	publisher = {{ACM}},
	year      = {2014}
}

@misc{lamarche:essential,
	TITLE = {{Proof Nets for Intuitionistic Linear Logic: Essential Nets}},
	AUTHOR = {Lamarche, Fran{\c c}ois},
	URL = {https://hal.inria.fr/inria-00347336},
	TYPE = {Research Report},
	YEAR = {2008},
	HAL_ID = {inria-00347336},
	HAL_VERSION = {v1}
}

@misc{hughes:conflict,
      title={Abstract p-time proof nets for MALL: Conflict nets}, 
      author={Dominic J. D. Hughes},
      year={2008},
      eprint={0801.2421},
      archivePrefix={arXiv},
      primaryClass={math.LO},
      url={https://arxiv.org/abs/0801.2421}, 
}

@article{str:MLL2, 
	title={Deep inference and expansion trees for second-order multiplicative linear logic}, 
	volume={29}, 
	DOI={10.1017/S0960129518000385}, 
	number={8}, 
	journal={Mathematical Structures in Computer Science},
	author={Strassburger, Lutz}, 
	year={2019}, 
	pages={1030–1060}
}

@inproceedings{hug:van:sliceLICS,
	author = {Hughes, Dominic J. D. and Van Glabbeek, Rob J.},
	title = {Proof Nets for Unit-free Multiplicative-Additive Linear Logic (Extended abstract)},
	year = {2003},
	isbn = {0769518842},
	publisher = {IEEE Computer Society},
	address = {USA},
	booktitle = {Proceedings of the 18th Annual IEEE Symposium on Logic in Computer Science},
	pages = {1},
	series = {LICS '03}
}

@article{hug:van:slice,
	author = {Hughes, Dominic J. D. and Van Glabbeek, Rob J.},
	title = {Proof nets for unit-free multiplicative-additive linear logic},
	year = {2005},
	issue_date = {October 2005},
	publisher = {Association for Computing Machinery},
	address = {New York, NY, USA},
	volume = {6},
	number = {4},
	issn = {1529-3785},
	url = {https://doi.org/10.1145/1094622.1094629},
	doi = {10.1145/1094622.1094629},
	journal = {ACM Trans. Comput. Logic},
	month = {oct},
	pages = {784–842},
	numpages = {59}
}

\end{document}